\documentclass[journal=jctcce,manuscript=article]{achemso}
\usepackage{amsmath}
\usepackage{amssymb}
\usepackage[version=4]{mhchem}
\usepackage{xcolor}
\usepackage{hyperref}
\usepackage{placeins}
\usepackage{braket}
\usepackage{subcaption}
\usepackage{caption}
\usepackage{mathrsfs}
\usepackage{extramarks}
\usepackage{geometry}
\usepackage{threeparttable}
\usepackage{array,multirow,makecell}
\usepackage{nicematrix}
\usepackage{tabularx}
\usepackage{booktabs}
\usepackage{comment}
\usepackage{xr}
\usepackage{placeins}
\author{Oskar Graulund Lentz Rasmussen}
\affiliation[SDU]{Department of Physics, Chemistry and Pharmacy, University of Southern Denmark, Campusvej~55, DK--5230 Odense M, Denmark}
\email{oskr@sdu.dk}

\author{Erik Kjellgren}
\affiliation[SDU]{Department of Physics, Chemistry and Pharmacy, University of Southern Denmark, Campusvej~55, DK--5230 Odense M, Denmark}

\author{Peter Reinholdt}
\affiliation[SDU]{Department of Physics, Chemistry and Pharmacy, University of Southern Denmark, Campusvej~55, DK--5230 Odense M, Denmark}

\author{Stephan P. A. Sauer}
\affiliation[KU]{Department of Chemistry, University of Copenhagen, Universitetsparken 5, DK--2100 Copenhagen Ø, Denmark}

\author{Sonia Coriani}
\affiliation[DTU]{Department of Chemistry, Technical University of Denmark, Kemitorvet Building 207, DK--2800 Kongens Lyngby, Denmark}

\author{Karl Michael Ziems}
\affiliation[UoS]{School of Chemistry, University of Southampton, Highfield, Southampton SO17 1BJ, UK}

\author{Jacob Kongsted}
\affiliation[SDU]{Department of Physics, Chemistry and Pharmacy, University of Southern Denmark, Campusvej~55, DK--5230 Odense M, Denmark}

\title{Cost-effective scalable quantum error mitigation for tiled Ans\"atze}

\begin{document}

\begin{abstract}
We introduce a cost-effective quantum error mitigation technique that builds upon the recent Ansatz-based gate and readout error mitigation method (M0). The technique, tiled M0, leverages the unique structure of tiled Ans\"atze (e.g., tUPS, QNP, hardware-efficient circuits) to apply a locality approximation to M0 that results in an exponential reduction in the QPU cost of the noise characterization. We validate the technique for molecular ground state energy calculations with the tUPS Ansatz on LiH, \ce{H2}, \ce{H2O}, butadiene, and benzene (4-12 qubits), demonstrating little to no loss in accuracy compared to M0 in noisy simulations. We also show the performance of the technique in quantum experiments, highlighting its potential use in near-term applications.
\end{abstract}
\section{Introduction}
Quantum computing is thought to offer solutions to classically intractable quantum simulation problems. Given the severity of noise on current noisy intermediate-scale quantum (NISQ) computers\cite{preskill2018quantum}, however, the touted benefits cannot yet be fully realized. Nonetheless, through the use of quantum error mitigation (QEM), which aims not to correct but to mitigate errors through classical data processing methods, it is possible to perform more accurate computational tasks with existing devices. Thus, even though hardware improvements (i.e., larger, faster, and less error-prone quantum computers) together with quantum error correction promise to unlock reliable quantum computing in the future\cite{shor1997, aharonov1997}, QEM will continue to play an important role in the current NISQ era and also beyond\cite{myths}.

Many QEM techniques, including the one we propose herein, require some characterization and knowledge of the noise on the quantum computer, which is then used to remove errors from measurement results through different schemes. In general, QEM techniques increase the sampling cost needed to estimate an observable with a desired accuracy and certainty. The sampling cost has been shown to increase exponentially with the circuit depth and gate error rates, regardless of the specifics of the QEM protocol\cite{Takagi_2023, Quek_2024}. This makes QEM useful only when the noise is at a manageable level or if the circuit is sufficiently shallow. A number of different QEM techniques have been developed for quantum computing such as zero-noise extrapolation (ZNE)\cite{Temme_2017, Li_2017, digital_zero_noise}, probabilistic error cancellation (PEC)\cite{Temme_2017, PhysRevX.8.031027}, readout error mitigation (REM)\cite{rem}, Clifford data regression (CDR)\cite{Czarnik_2021, PRXQuantum.2.040330, Czarnik2025improvingefficiency, zhao2025}, tensor-network error mitigation (TEM)\cite{filippov2023}, and Ansatz-based gate and readout error mitigation (M0)\cite{ziems}. An introduction to the field of QEM, as well as a review of some of the aforementioned techniques, is given in Ref.~\citenum{Cai_2023}.

The main idea behind our proposed QEM technique is to utilize the layered structure of tiled Ans\"atze such as the tiled unitary product state (tUPS) Ansatz\cite{tups} to characterize gate and readout noise efficiently. We focus here on the tUPS Ansatz, but in principle the technique also works for other tiled Ans\"atze such as quantum-number-preserving\cite{Anselmetti2021-qg}  and hardware-efficient Ans\"atze with tiled structures\cite{Berthusen2022-wa,Sun2023-da,Tepaske2023-yp,Ayeni2025-ip,Mihalikova2025-st,Uvarov2020-wt}. Our technique builds upon the recently proposed M0 QEM method by Ziems, Kjellgren, et al.,\cite{ziems} where the authors extend the REM technique by not only measuring the qubits during the noise characterization step, but by also including the Ansatz with all variational parameters set to zero. 
This approach of using a zero-parameterized Ansatz as part of noise characterization has found use in other QEM methods\cite{motta2023quantum, lolur2023reference, zou2025multireference}. For Ans\"atze such as tUPS, the zero-parameterized circuit is an identity operation on a noiseless device. In the presence of noise, however, the inclusion of the Ansatz will have the effect of encoding gate noise in addition to readout noise in the confusion matrix. The M0 technique has been used in hardware experiments\cite{ziems,Jensen2025-dr} and simulated noise studies\cite{peter}, but it requires the construction of a single fully correlated confusion matrix. This brings about an exponential quantum processing unit (QPU) measurement cost just from the dimensions of the matrix alone, thereby limiting its application to small systems. 

In our proposed technique, which we refer to as \textit{tiled M0}, we combine M0 with a locality approximation that exploits the repeating structure of tiled Ans\"atze. The approximation is similar to the tensor product approach suggested for REM in Ref.~\citenum{rem}, and it results in an exponential reduction in the QPU noise characterization cost compared to M0. The reduction is achieved by approximating the full-system confusion matrix through a number of smaller matrices - one for each tile in the Ansatz - and, in particular, it is the number of distinct noise characterization circuits which need to be executed that are reduced. The assumption is that correlated noise between qubits that share a tile in the Ansatz will be more dominant for QPUs with static layouts, similar to the idea of qubit neighborhoods in Ref.~\citenum{geller2021toward}.

We show the performance of tiled M0 for ground state energy calculations with the tUPS Ansatz on different molecules in quantum experiments using IBM quantum computers and in noisy simulations. We look at LiH and \ce{H2} with $(2,2)$ active spaces and up to four layers in the tUPS Ansatz, butadiene and \ce{H2O} with $(4,4)$ active spaces and one layer in the Ansatz, and benzene with a $(6,6)$ active space and one layer in the Ansatz.

\section{Background}\label{sec:theory}
\subsection{Tiled unitary product state Ansatz}
\begin{figure}
\centering
  \centering
  \includegraphics[width=0.75\linewidth]{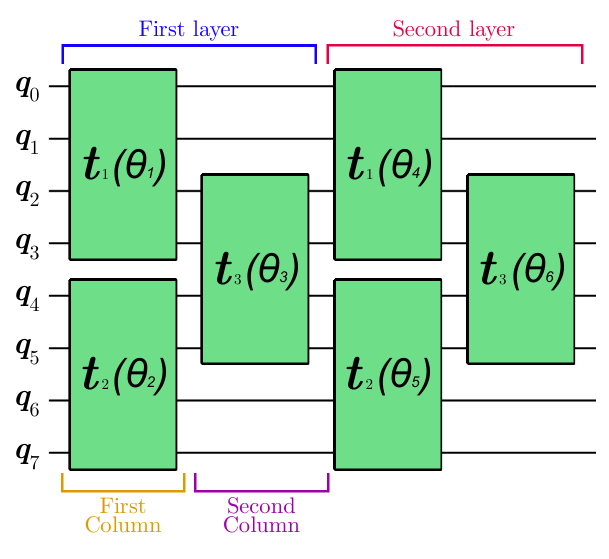}
\caption{The tUPS Ansatz with two layers for an $8$-qubit system. Each tile (labelled $t_i$) consists of gates that correspond to single and double electronic excitations. Three variational parameters are associated with every tile.}
  \label{fig:tups}
\end{figure}

The tUPS Ansatz introduced by Burton in Ref.~\citenum{tups} is a commonly employed tiled Ansatz in quantum computational chemistry (see Fig.~\ref{fig:tups} for an illustration of the structure). A tile in tUPS is a unitary operator that causes electronic transitions between four spin orbitals using three variational parameters, which ensures that any state can be reached in the subset\cite{tups, burton2023exact}. Each layer consists of two columns of tiles (one if the number of qubits is less than $6$). The tiles in the second column of a layer in tUPS are offset compared to the tiles in the first column, which is necessary to ensure that any electronic configuration can be reached, given enough layers\cite{evangelista2019exact}. Generally, the more layers that are included in the Ansatz, the greater the expressivity. Following Ref.~\citenum{tups}, the unitary operator for a tile can be expressed in terms of the fermionic excitation operators $\hat{\kappa}_{pq}^{(1)}$ and $\hat{\kappa}_{pq}^{(2)}$ as
\begin{equation}
\hat{U}^{(m)}_{pq} = \text{exp}\left(\theta^{(m)}_{pq,1} \hat{\kappa}^{(1)}_{pq} \right) \text{exp}\left(\theta^{(m)}_{pq,2} \hat{\kappa}^{(2)}_{pq} \right) \text{exp}\left(\theta^{(m)}_{pq,3} \hat{\kappa}^{(1)}_{pq} \right),
\label{eq:tile}
\end{equation}
where $p$ and $q$ are spatial orbital indices and $m$ refers to the layer number (see Fig. \ref{fig:tile} in the supporting information (SI) for a quantum circuit diagram). The tUPS Ansatz for an $n$-qubit system is then
\begin{equation}
\hat{U}_{\mathrm{tUPS}}(\theta) = \prod^L_{m=1}\left(\prod^R_{p=1}\hat{U}^{(m)}_{2p+1,2p}\prod^S_{p=1}\hat{U}^{(m)}_{2p,2p-1}\right),
\end{equation}
where $L$ is the number of layers, and $S =\mathrm{floor}(\frac{n}{4})$ and $R = \mathrm{floor}(\frac{n-2}{4})$ are the number of tiles in the first and second column, respectively. The tile numbering $t_i$ as shown in Fig.~\ref{fig:tups} relates to the tUPS Ansatz numbering in terms of $p$ and $q$ as follows (with one-based indexing for $p$ and $q$):
\begin{align}
    p &= \begin{cases}
    2 i -1, & \text{if $i\leq S$} \quad \text{(first columns)};\\
    2(i-S), & \text{otherwise} \quad \text{(second columns)}.
  \end{cases} \\
    q &= p + 1
\end{align}
These in turn relate to the qubit number on which a tile starts by $q_\mathrm{num} = 2(p-1)$. For clarity and ease of presentation, in the following we will only refer to the tile numbering, $t_i$, and refer to the above for connecting a specific tile to its unitary and circuit implementation. 

\subsection{Readout error mitigation}
Our proposed QEM technique (\textit{vide infra}) is based on readout error mitigation (REM). REM was introduced by Bravyi et al.\cite{rem} to correct readout noise under the assumption that the noise can be modelled by a stochastic matrix, $\mathbf{A}^{\mathrm{REM}}$, called an assignment, transition or confusion matrix. The entry $(\mathbf{A}^{\mathrm{REM}})_{yx}$, which is referred to as a transition probability, gives the probability of measuring the bit string $\ket{y}$ when the system was prepared in the basis state $\ket{x}$:
\begin{align}
    (\mathbf{A}^{\mathrm{REM}})_{yx} = \mathrm{Pr}(\ket{y} \, | \, \ket{x}).
    \label{eq:rem_Pr}
\end{align}
To construct the full confusion matrix, every possible bitstring $\ket{x} \in \left\{\ket{b_i^\mathrm{REM}}\right\}$ needs to be prepared and measured. These are defined as
\begin{equation}
    \ket{b_i^\mathrm{REM}} = \hat{X}_i\ket{0}^{\otimes n}
    \label{prep_rem}
\end{equation}
where $\hat{X}_i$ is an operator with the $X$-gates necessary to create any bit string $\ket{b_i}$.
When the state $\ket{x} \in \left\{\ket{b_i^\mathrm{REM}}\right\}$ has been prepared according to Eq.~\eqref{prep_rem}, the entry $(\mathbf{A}^{\mathrm{REM}})_{yx}$ is estimated from a finite number of $K$ measurement results, $\ket{s_1},...,\ket{s_K}$, as
\begin{equation}
    (\mathbf{A}^{\mathrm{REM}})_{yx} \approx K^{-1} \sum_{i = 1}^{K} \braket{s_i|y}.
    \label{rem_est}
\end{equation}
The idea proposed with the M0 technique \cite{ziems} is to include not only $X$-gates when the basis states are prepared, but the full Ansatz $\hat{U}$ with all variational parameters set to $0$, i.e.
\begin{equation}
   \ket{b_i^\mathrm{M0}} = \hat{U}(0)  \hat{X}_i\ket{0}^{\otimes n}~,
   \label{prep_ziems}
\end{equation}
and use this to construct the M0 confusion matrix, $\mathbf{A}^{\mathrm{M0}}$, in analogy with $\mathbf{A}^{\mathrm{REM}}$, with entries defined by expressions analogous to Eqs.~\eqref{eq:rem_Pr} and \eqref{rem_est}. In the absence of hardware noise, $\hat{U}(0)$ is the identity operator for Ans\"atze like tUPS, and its inclusion has no effect. On noisy devices, however, the errors associated with the gates in $\hat{U}$ will influence the transition probabilities, thereby encoding gate noise in the confusion matrix. In the remainder of the text, we refer to confusion matrices $\mathbf{A}$ constructed from a general basis state preparation circuit and omit qualifiers such as $\mathrm{REM}$ and $\mathrm{M0}$ unless there is need for a distinction.

\subsection{Expectation value estimation} \label{sec:expect_estimation}
The error characterized in $\mathbf{A}$ is applied to a subsequent noisy simulation or quantum experiment as follows: Given a set of $N$ noisy measurement results, $\ket{r_1},..., \ket{r_N}$, the error mitigated outcomes are obtained by multiplying with the inverse of the confusion matrix, i.e.\ $\mathbf{A}^{-1}\ket{r_1}, ..., \mathbf{A}^{-1}\ket{r_N}$. 
In reality, defining the raw (unmitigated) probability vector  $\ket{p_\mathrm{raw}} = N^{-1} \sum_{i = 1}^N\ket{r_i}$, we can write the error mitigated results as 
\begin{equation}
    \ket{p_\mathrm{mitigated}}=\mathbf{A}^{-1} \ket{p_\mathrm{raw}}
    \label{p_mitigated}
\end{equation}
Thus, an estimate of the expectation value of an operator $\hat{O}$ which is diagonal in the computational basis is obtained as
\begin{equation}
    \bar{O} = \sum_{x\in \{0, 1\}^n} O_x \braket{x|\mathbf{A}^{-1}|p_\mathrm{raw}}
\end{equation}
where $n$ is the number of qubits and $O_x$ is the eigenvalue corresponding to the computational basis state $\ket{x}$.

\subsection{Tiled error mitigation}
The major drawback of REM and M0 is the exponential scaling in the QPU processing time needed for the noise characterization which arises from the scaling of the bitstring space (see Eqs.~\eqref{prep_rem} and \eqref{prep_ziems}). This is an extra cost that is unrelated to the circuit error rates which inherently bring about their own exponential overhead\cite{Takagi_2023, Quek_2024}.
To alleviate the bitstring space scaling problem, we combine M0 with the concept of noise locality from the REM tensor product model and related works\cite{rem, geller2021toward}. In the tensor product model, noise is assumed to act locally on individual qubits and $\mathbf{A}$ is approximated as a tensor product of the $2 \times 2$ confusion matrices for all qubits, i.e.
\begin{equation}
    \mathbf{A} \approx \mathbf{A}_{q_{n-1}} \otimes \cdots  \otimes \mathbf{A}_{q_0}
    \label{tensor_method}
\end{equation}
where $\mathbf{A}_{q_i}$ is the confusion matrix for qubit $i$. The approximation results in an exponential saving in the QPU cost needed to construct $\mathbf{A}$. In tiled M0, we assume noise locality for qubits that share a common tile in the Ansatz, and the crux of our technique is that we construct a confusion matrix for every tile in the first layer of the Ansatz. For ease of explanation, assume in the following that there is no readout but only gate noise (we deal with readout noise afterwards). We explain the underlying ideas and advantages of the technique in the context of the tUPS Ansatz for concreteness, but note that it is applicable to other tiled Ans\"atze.

Let $t_i$ refer to tile number $i$ in the tUPS Ansatz (see Fig.~\ref{fig:tups} for the ordering) and let $\mathbf{A}_{t_i}$ be the corresponding tile confusion matrix (we explain what quantum circuits we execute to determine $\mathbf{A}_{t_i}$ momentarily). We assume that tile confusion matrices are the same across layers, i.e. $\mathbf{A}_{t_i} \approx \mathbf{A}_{t_{(i+mT)}}$ where $T$ is the number of tiles in the first layer of the Ansatz and $m \in \mathbb{N}$. We refer to this as the layer approximation, and we approximate the full $n$-qubit confusion matrix for $n \geq 4$ and $n$ even as
\begin{align}
    \mathbf{A} &\approx \begin{cases}
    \Big[(\mathbf{I}^{\otimes 2} \otimes \mathbf{A}_{t_T} \otimes \cdots \otimes \mathbf{A}_{t_{S+1}} \otimes \mathbf{I}^{\otimes 2})(\mathbf{A}_{t_S} \otimes \cdots \otimes \mathbf{A}_{t_1}) \Big]^L, & \text{if $4 \mid n$}~;\\
    \Big[( \mathbf{A}_{t_T} \otimes \cdots \otimes \mathbf{A}_{t_{S+1}} \otimes \mathbf{I}^{\otimes 2})(\mathbf{I}^{\otimes 2} \otimes \mathbf{A}_{t_S} \otimes \cdots \otimes \mathbf{A}_{t_1}) \Big]^L, & \text{otherwise};
  \end{cases}
  \label{tile_approx}
\end{align}
where $S = \text{floor}(\frac{n}{4})$ is the number of tiles in the first column of the Ansatz as defined previously and $T = \mathrm{floor}(\frac{n}{2}) - 1$. $T$ scales linearly with the number of qubits, so we only need to construct a linearly scaling number of tile confusion matrices to approximate $\mathbf{A}$.

The confusion matrix for a given tile is determined by including only the part of the Ansatz that belongs to that tile as given by Eq.~\eqref{eq:tile} (note that this reduces to the identity when the variational parameters are all zero). Since each tile has four qubits in the tUPS Ansatz, we need to prepare only $16$ different computational basis states on the QPU for every tile, i.e., $16$ different circuits, to determine the transition probabilities (with a sufficient amount of shots for each). Moreover, since tiles that lie in the same column in the tiled Ansatz have no overlapping qubits, we can prepare and measure the computational basis states for each of those tiles in parallel. For the example in Fig.~\ref{fig:tups}, this means that we determine $\mathbf{A}_{t_1}$ and $\mathbf{A}_{t_2}$ from the same probability vectors. Since there are two columns in a layer (for systems comprised of $6$ or more qubits), and because of the layer approximation, we need to execute only $32$ different quantum circuits to construct all tile confusion matrices for arbitrary layers and depth. We now explain how readout noise fits in this picture.

\subsection{Preventing the overcorrection of readout noise}
When we determine the transition probabilities as described in the previous section, the tile confusion matrices will encode not only gate noise but also readout noise. If at any point we have two or more tile confusion matrices in Eq.~\eqref{tile_approx} that share any qubits, we will correct the readout noise for those qubits more than once. This is expected to introduce a bias in the error-mitigated results, which will become worse the more tiles and layers we have. To handle this problem, we work with the assumption that any confusion matrix can be written approximately as the product of a readout part and a gate part
\begin{equation}
    \mathbf{A}_{t_i} \approx  \mathbf{A}_{t_i}^\mathrm{REM} \mathbf{A}_{t_i}^\mathrm{gate}~~.
    \label{gate_readout_approx}
\end{equation}
In other words, we assume that gate and readout noise are completely independent of each other, and we refer to Eq.~\eqref{gate_readout_approx} as the separability approximation. Since $\mathbf{A}_{t_i}^\mathrm{REM}$ can be determined separately for the qubits in every tile using the traditional REM approach, we can remove the readout part from $\mathbf{A}_{t_i}$, thereby obtaining the pure gate noise confusion matrix
\begin{equation}
    \mathbf{A}_{t_i}^\mathrm{gate} \approx (\mathbf{A}_{t_i}^\mathrm{REM})^{-1} \mathbf{A}_{t_i}~~.
\end{equation}
In the context of tUPS, this yields the full-system gate error confusion matrix similar to Eq.~\eqref{tile_approx} as
\begin{align}
    \mathbf{A}^\mathrm{gate} &\approx \begin{cases}
    \Big[(\mathbf{I}^{\otimes 2} \otimes \mathbf{A}_{t_T}^\mathrm{gate} \otimes \cdots 
    \otimes \mathbf{A}_{t_{S+1}}^\mathrm{gate} \otimes \mathbf{I}^{\otimes 2})(\mathbf{A}_{t_S}^\mathrm{gate} \otimes \cdots \otimes \mathbf{A}_{t_1}^\mathrm{gate}) \Big]^L, & \text{if $4 \mid n$}~;\\
    \Big[( \mathbf{A}_{t_T}^\mathrm{gate} \otimes \cdots \otimes \mathbf{A}_{t_{S+1}}^\mathrm{gate} \otimes \mathbf{I}^{\otimes 2})(\mathbf{I}^{\otimes 2} \otimes \mathbf{A}_{t_S}^\mathrm{gate} \otimes \cdots \otimes \mathbf{A}_{t_1}^\mathrm{gate}) \Big]^L, & \text{otherwise},
  \end{cases}
  \label{a_gate}
\end{align}
Under the working assumptions, this matrix encodes all the gate noise that would be captured in an M0 experiment, but none of the readout noise. Note that we need to execute $64$ different quantum circuits in total to construct the gate error tile confusion matrices in tUPS 
- a number that is independent of the system size (for systems comprised of $6$ or more qubits).
The full-system readout confusion matrix $\mathbf{A}^{\mathrm{REM}}$ can be approximated as
\begin{align}
    \mathbf{A}^\mathrm{REM} & \approx \begin{cases}
    \mathbf{A}^{\mathrm{REM}}_{t_S} \otimes \cdots \otimes \mathbf{A}_{t_1}^{\mathrm{REM}}, & \text{if $4 \mid n$}~;\\
    \mathbf{A}_{q_{n-1},q_{n-2}}^{\mathrm{REM}}  \otimes \mathbf{A}^{\mathrm{REM}}_{t_S} \otimes \cdots \otimes \mathbf{A}_{t_1}^{\mathrm{REM}}, & \text{otherwise};
  \end{cases}
  \label{adv_rem}
\end{align}
where $\mathbf{A}^{\mathrm{REM}}_{q_{n-1}, q_{n - 2}}$ is the $4 \times 4$ two-qubit readout confusion matrix for the last two qubits. Thus, the readout error characterization uses the first column REM confusion matrices $\mathbf{A}^{\mathrm{REM}}_{t_i}$ that we construct to determine $\mathbf{A}_{t_i}^\mathrm{gate}$ anyways, and therefore comes at no additional cost in the case of $4 \mid n$ or with only $4$ additional circuits to measure otherwise. This means that the complete noise characterization costs at most $(64 + 4) \times K$ quantum circuit measurements, where $K$ is the shot count (we explain how we chose $K$ in this work in Sec.~\ref{sec:computational_details}).

To summarize, tiled M0 approximates the full-system M0 confusion matrix $\mathbf{A}^{\mathrm{M0}}$ as
\begin{equation}
    \mathbf{A}^{\mathrm{M0}} \approx \mathbf{A}^\mathrm{REM} \mathbf{A}^\mathrm{gate}
    \label{eq:m0_approximation}
\end{equation}
with the matrices on the right-hand side given by Eqs.~\eqref{a_gate} and \eqref{adv_rem}. However, the full-system matrix need not be constructed explicitly in tiled M0; rather, each inverse tile confusion matrix can be acted in turn on the raw probability vectors, providing the same combined effect as the full matrix. While tiled M0 removes the exponential scaling in the confusion matrix, we note that it still requires explicit construction of the probability vectors. Thus, classical post-processing remains exponentially costly.

\subsection{Underlying assumptions}
We briefly summarize the underlying assumptions of tiled M0. 
First, in tiled M0, hardware errors are treated in an entirely classical manner and modelled as stochastic processes. Each tile in the zero-parameterized Ansatz is viewed as inducing classical transitions between computational basis states via confusion matrices, independently of other tiles. If the transitions of two qubits that do not share a common tile in the Ansatz are correlated, tiled M0 will fail to capture this information. As such, tiled M0 is expected to perform poorly on a device dominated by highly non-local noise, e.g. crosstalk errors. 
Second, we assume readout and gate error separability in Eq.~\eqref{gate_readout_approx}. This assumption is motivated by the well-established effectiveness of REM. If the readout noise was highly dependent on the specifics of the circuit being executed, then such circuit-independent readout calibration would no longer be reliable, and the validity of both REM and the separability approximation would be compromised.
Third is the layer approximation, i.e. that the same tiles in different layers induce the same transitions between computational basis states. The validity of this approximation hinges on the validity of the classical treatment of the noise altogether, which we investigate numerically and via quantum experiment in this work.

In addition, tiled M0 shares the noise stability assumption of most QEM methods: if the noise changes rapidly during or after noise characterization, the confusion matrices will not be representative of the noise at the time of expectation value measurement. In Sec.~\ref{noise}, we discuss this further.

Finally, tiled M0 assumes that noise characterized at zero variational parameters, $\vec{\theta} = 0$, remains informative for the parameterized circuits used in expectation value estimation where $\vec{\theta} \neq 0$. The effect of nonzero variational parameters on the accuracy of both M0 and tiled M0 as well as related works that share this assumption\cite{motta2023quantum, lolur2023reference, zou2025multireference} is not yet fully understood. However, because the dominant gate errors in the circuits considered here are expected to arise primarily from two-qubit gates, we expect parameter-dependent noise associated with single-qubit rotations to make a comparatively small contribution to the overall error. The same can be said for single-qubit basis state rotation gates which are not included as part of the noise characterization circuits in tiled M0. When we refer to tiled M0 as mitigating gate errors, it should be understood that this does not include basis state rotation gates.

\subsection{Noise strength and confusion matrix condition numbers\label{sec:overhead}}
For the tensor product model, Bravyi et al. show in Ref.~\citenum{rem} that the number of shots needed to estimate the expectation value of an observable with a desired accuracy and certainty scales exponentially with the circuit noise strength,
\begin{equation}
    \gamma = \sum_{i = 0}^{n - 1} \mathrm {max}\Big[(\mathbf{A}_{q_i})_{10}, (\mathbf{A}_{q_i})_{01}\Big]~~,
    \label{gamma_overhead}
\end{equation}
which is governed by the off-diagonal entries over all $\mathbf{A}_{q_i}$ in Eq.~\eqref{tensor_method}. This extra sampling overhead is referred to as the error mitigation overhead. Since the confusion matrices $\mathbf{A}_{q_i}$ can be approximated at very little QPU cost (we use $2 \times 15,000$ shots independent of the system size and depth), they can be used as a cheap way to estimate the overhead and noise severity for a given circuit and backend at a given point in time. 



The matrix condition number, $\kappa$, is another useful measure of the noise severity. For a matrix $\mathbf{A}$ with maximal and minimal singular values $\sigma_{\text{max}}(\mathbf{A})$ and $\sigma_{\text{min}}(\mathbf{A})$, the condition number is defined as
\begin{equation}
    \kappa(\mathbf{A}) = \frac{\sigma_{\text{max}}(\mathbf{A})}{\sigma_{\text{min}}(\mathbf{A})}~~.
\end{equation}
A large condition number of a confusion matrix is indicative of substantial noise. In the ideal zero-noise limit, all confusion matrices in tiled M0 reduce to identity matrices and therefore have condition number one. In the opposite limit of complete randomness, all entries in the confusion matrices become identical, so that the matrices are singular. These ill-conditioned matrices are numerically unstable under inversion: small perturbations in their entries can lead to large changes in the entries of the inverse. Consequently, the expectation value estimation procedure described in Sec.~\ref{sec:expect_estimation} becomes increasingly susceptible to error as the condition numbers of the relevant confusion matrices grow.
Hence, the success of tiled M0 can be seen to be largely dependent on the condition numbers. A useful property of the condition number is $\kappa(\mathbf{A}\otimes \mathbf{B}) = \kappa(\mathbf{A}) \kappa(\mathbf{B})$ which follows from $||\mathbf{A} \otimes \mathbf{B}|| = ||\mathbf{A}|| \, ||\mathbf{B}||$ where $||  \cdot ||$ denotes the spectral norm\cite{lancaster1972norms}. Condition numbers for the full-system confusion matrices used in the quantum experiments in this work range between $1.29$ and $19.2$ apart from a few outliers with values up to the order of $10^5$ (more details in Sec.~\ref{results}).

\section{Computational Details}\label{sec:computational_details}
We performed ground state energy calculations with the tUPS Ansatz on equilibrium geometries of: 1) \ce{LiH} and \ce{H2} with $(2,2)$ active spaces and up to four layers in the tUPS Ansatz; 2) \ce{H2O} and butadiene with $(4,4)$ active spaces and one layer in the Ansatz and 3) benzene with a $(6,6)$ active space and one layer in the Ansatz. With the severity of noise encountered during the quantum experiments, it became infeasible to go beyond one layer for the larger systems, and we deemed it sufficient to study LiH and \ce{H2} up to four layers.

\subsection{Workflow overview}
The fermionic Hamiltonians and quantum circuits for all molecular systems were obtained with the Python library SlowQuant\cite{slowquant} using the STO-3G basis set\cite{Hehre1969-il}. Qubit Hamiltonians were obtained from the fermionic Hamiltonians with the Jordan-Wigner transformation using the Qiskit Nature library\cite{qiskit, qiskit_nature}. Reference energies and optimal parameters for the tUPS Ans{\"a}tze were calculated classically with SlowQuant's ideal state vector simulator. PySCF\cite{pyscf} was used for the initial Hartree-Fock calculations. Qiskit and Qiskit Aer were utilized for all device noise simulations, and all quantum experiments were performed on IBM quantum computers through the IBM Quantum Platform, Qiskit, and the Qiskit IBM Runtime library\cite{qiskit}. For the hardware experiments, the \texttt{ibm\_fez} and \texttt{ibm\_marrakesh} backends were used. The device noise simulations adopted noise models imported from those same backends and included transpilation (with a fixed seed and respecting the connectivity map of the backend). Perfect-pairing was used for the input states for all energy calculations \cite{tups}.

\subsection{Overhead screening}
We performed error mitigation overhead screening before carrying out most of the quantum experiments in this work. This was done to prevent the waste of QPU resources in the case of severe noise where error mitigation is infeasible. The screening involved approximating all $\mathbf{A}_{q_i}$ matrices in Eq.~\eqref{tensor_method} and determining the noise strength $\gamma$ through Eq.~\eqref{gamma_overhead}. The $\mathbf{A}_{q_i}$ matrices were determined exactly as in the REM tensor product model, the only difference being that the calibration circuits included the full zero-parameterized Ansatz for the molecule in the question. For each screening, we used $15,000$ shots for each of the two calibration circuits needed to determine the $\mathbf{A}_{q_i}$ matrices. In most cases, we deemed the overhead to be at a reasonable level, and we immediately proceeded with the main calculation following screening. However, in a few cases the noise was found to be unacceptably high and we postponed the calculation. The decision to postpone or not was based on how close the noise strengths were compared to values calculated on a noisy simulator which ranged from $\gamma_{\text{sim}} = 0.3$ for LiH and \ce{H2} at one layer to $\gamma_{\text{sim}} = 1.4$ for benzene. No strict cutoff values were chosen, but if the noise strength was greater than $2\gamma_{\text{sim}}$, we would generally choose to postpone.

\subsection{Noise characterization}
We executed all quantum circuits needed for the construction of any confusion matrix with a shot count that determines the transition probabilities with an accuracy of at least $10^{-2}$ with a probability of at least $90 \%$. The shot count can be determined from Hoeffding's inequality, which provides an upper bound on the probability that a sample mean value, $\bar{x}$, differs more than $\delta$ from the corresponding population mean value, $\mu$,
\begin{equation}
    P(|\bar{x} - \mu| \geq \delta) \leq 2\exp\bigg(-\frac{2K^2\delta^2}{\sum_i^K(b_i - a_i)^2}\bigg) \label{eq:hoeff}
\end{equation}
where $\bar{x}$ is the mean of $K$ measurement results $x_1, ..., x_k$, and $a_i$ and $b_i$ are lower and upper bounds for the measurement result $x_i$, i.e. $a_i \leq x_i \leq b_i$. For a given confusion matrix entry, $\mu$ is the exact transition probability as given by Eq.~\eqref{eq:rem_Pr} or its tiled M0 analog while $\bar{x}$ is the approximated transition probability as given by Eq.~\eqref{rem_est}. The lower and upper bounds are $a_i = 0$ and $b_i = 1$ for all $i$ for the transition probabilities, and the necessary shot count for $\delta = 10^{-2}$ and an upper bound of $10\%$ is $14,979$ for each computational basis state. For the butadiene, \ce{H2O} and benzene $(4,4)$ and $(6,6)$ systems, we used $64 \times 14,979 = 958,656$ shots in total to construct all tile confusion matrices in Eqs.~\eqref{a_gate} and \eqref{adv_rem} which corresponded to approximately $5$ minutes of QPU time being spent characterizing the noise (on \texttt{ibm\_fez} and \texttt{ibm\_marrakesh}). For the \ce{H2} and \ce{LiH} $(2,2)$ systems, we used $479,328$ shots in total to construct the matrices. An improved approach could be to allocate more shots to circuits with more qubits and greater depth (i.e. noisier circuits) to avoid excessive error propagation upon confusion matrix inversion. Whether this will be advantageous in practice in the presence of factors such as noise drift is part of future works.

To preserve the structure of the noise characterization circuits during transpilation with Qiskit, we left all variational parameters unassigned during transpilation. This prevented Qiskit from optimizing away gates that became trivial only after parameter assignment, while still permitting other transpiler optimizations. The variational parameters were subsequently assigned the value $0$, yielding the desired characterization circuits.
Basis state rotation gates were not included in any noise characterization circuit. The subsequent energy measurement circuits that were executed (more details in Sec.  \ref{sec:ham_meas} below) used the exact same tile circuits, i.e. identical gates and physical qubits, as noise characterization except for the variational parameters and possible basis state rotation gates.

\subsection{Energy measurements} \label{sec:ham_meas}
For measuring the electronic energy, we employed a heuristic minimum clique cover algorithm to group the Pauli strings of the qubit Hamiltonian based on qubit-wise commutativity (QWC) using the largest-first approach\cite{Verteletskyi2020-nn,Gokhale2020-lw}. Again, we used Hoeffding's inequality (see Eq.~\eqref{eq:hoeff}) to determine the number of shots that were required for each clique to estimate the energy with a desired accuracy and certainty. This necessitated an estimate of the error mitigation overhead that we obtained through the procedure described in Sec.~\ref{sec:overhead}.

The approach of using Hoeffding's inequality is approximate, but it allows us to allocate more shots to cliques of Pauli strings that have large absolute weights. To briefly illustrate how the weights play in, consider an operator $\hat{O} = w_1 \hat{O}_1 +w_2 \hat{O}_2$ where $\hat{O}_1$ and $\hat{O}_2$ are Pauli strings that commute qubit-wise, and $w_1$ and $w_2$ are the corresponding weights. Assume that $\hat{O}_1$ and $\hat{O}_2$ are diagonal in the computational basis, and that we want to determine the expectation value of $\hat{O}$ for an arbitrary state - without error mitigation for simplicity. Each measurement result will yield one of the following values for $\hat{O}$: $w_1 + w_2$, $w_1 - w_2$, $-w_1 + w_2$, or $-w_1 - w_2$. The lower and upper bounds for the measurement results are $-(\left|w_1 \right| + \left|w_2 \right|)$ and $\left|w_1 \right| + \left|w_2 \right|$, respectively, so the denominator in the exponential in Eq.~\eqref{eq:hoeff} is $4K(\left|w_1 \right| + \left|w_2 \right|)^2$.  In general, the denominator will be $4K(\sum_j\left| w_j \right|)^2$, where the sum is over all Pauli strings in the clique. The larger the sum, the more shots will be needed to estimate the expectation value with a fixed accuracy and certainty.

The number of energy measurement shots used in all experiments are shown in Table~\ref{h_shots_tabel}. We repeated the experiments up to five times for each molecular system. The same number of shots was used in both quantum experiments and noisy simulations on the same molecule. Approximations of the shot noise uncertainties are shown in Table~\ref{shot_noise_tabel}.


\begin{table}[t]
\begin{threeparttable}
\centering\renewcommand\cellalign{lc}
\setcellgapes{3pt}\makegapedcells
\caption{The number of shots used to measure the energy for all molecular systems for each repeated calculation. Details about the qubit Hamiltonians are given in Table~\ref{shots_tabel}.}
\centering\renewcommand\cellalign{lcc}
\setcellgapes{3pt}\makegapedcells
\begin{tabularx}{\linewidth}{l|XXXXXX}
\hline
Molecule & LiH  &  \ce{H2} &  Butadiene & \ce{H2O} & Benzene \\  \hline  \hline
One layer & $117,696$ & $196,179$ & $2,467,569$ & $14,790,613$ & $25,203,951$ \\
Two layers & $221,964$ & $369,856$ & - & - & - \\
Three layers & $366,452$ & $610,631$ & - & - & - \\
Four layers & $758,861$ & $1,264,562$ & - & - & - \\
\bottomrule
\end{tabularx}
 \label{h_shots_tabel}
\end{threeparttable}
\end{table}

\begin{table}[t]
\begin{threeparttable}
\centering\renewcommand\cellalign{lc}
\setcellgapes{3pt}\makegapedcells
\caption{Approximated shot noise uncertainties for the energy calculations with the shot counts in Table~\ref{h_shots_tabel}. The reported uncertainties are standard deviations in units of hartree calculated from samples of $10^3$ energies, each obtained from a simulator without simulated device noise using the same number and distribution of shots among QWC-groups as for the noisy experiments.}
\centering\renewcommand\cellalign{lcc}
\setcellgapes{3pt}\makegapedcells
\begin{tabularx}{\linewidth}{l|XXXXX}
\hline
Molecule & LiH  &  \ce{H2} & Butadiene & \ce{H2O} & Benzene  \\  \hline \hline
One layer & $0.0031$ & $0.0073$ & $0.0049$ & $0.0033$ & $0.0051$ \\ 
Two layers & $0.0024$ & $0.0055$ & - & - & -\\ 
Three layers & $0.0019$ & $0.0043$ & - & - & - \\
Four layers & $0.0013$ & $0.0030$ & - & - & - \\
\bottomrule
\end{tabularx}
\label{shot_noise_tabel}
\end{threeparttable}
\end{table}

\subsection{System details}
Details about the qubit Hamiltonians for all molecular systems are shown in Table~\ref{shots_tabel} with information on the number of Pauli strings, the number of QWC cliques, and the L1 norms of the Pauli string weights (excluding the weight of the all-identity Pauli string). The number of times the energy calculation experiments were repeated, the total number of shots used (including both noise characterization and energy measurement), and the QPU processing times are also summarized. The number of shots that we used to construct the M0 confusion matrix for an $n$-qubit system was $14,979 \times 2^n$, and those shots are not included in the table. Note that the shot counts that are shown for $\ce{H2}$ and $\ce{LiH}$ are the total combined shots that were used for all the layer calculations, while the shot counts for the other molecules are for experiments with just one layer.

\begin{table}[t]
\begin{threeparttable}
\centering\renewcommand\cellalign{lc}
\setcellgapes{3pt}\makegapedcells
\caption{Details about the qubit Hamiltonians for the tested molecular systems, the number of times the energy calculations were repeated, the total number of shots, and the QPU processing times.}
\centering\renewcommand\cellalign{lcc}
\setcellgapes{3pt}\makegapedcells
\begin{tabularx}{\linewidth}{l|XXXXXX}
\hline
Molecule & LiH  &  \ce{H2}   &  Butadiene & \ce{H2O} & Benzene \\  \hline  \hline
Number of qubits & 4 & 4 & 8 & 8 & 12 \\
Pauli strings & 27 &  15 &  185 & 361 & 555 \\ 
Cliques &  9  & 5 &  46 & 73 & 132 \\ 
L1 norm & 1.47 &  1.89 &  5.06 & 9.44 & 9.15 \\
Shots, total\tnote{{$*$}}  & 1,944,301\tnote{{$\dagger$}} &  2,920,556\tnote{{$\dagger$}} & 3,426,225 & 15,749,269& 26,162,607\\
Repeats\tnote{{$\ddagger$}} & $(5; 5)$ &  $(5; 5)$ & $(5; 5)$ & $(3; 5)$ & $(3; 5)$ \\
QPU time, total\tnote{{$*$}} & $\sim 10 \, \mathrm{min}$ &  $\sim 15 \, \mathrm{min}$ &  $\sim 20 \, \mathrm{min}$ & $\sim 1.3 \, \mathrm{hrs}$ & $\sim 2.1 \, \mathrm{hrs}$  \\
QPU time, tiled M0 & $\sim 2.5 \, \mathrm{min}$&  $\sim 2.5 \, \mathrm{min}$ &  $\sim 5 \, \mathrm{min}$ & $\sim 5 \, \mathrm{min}$ & $\sim 5 \, \mathrm{min}$ \\

\bottomrule
\end{tabularx}
\begin{tablenotes}\footnotesize
\item[{$*$}] Per calculation (including both tiled M0 noise characterization and energy measurement).
\item[{$\dagger$}] The shot counts for \ce{H2} and \ce{LiH} are the total combined shots that were used for all the layer calculations from $1$ to $4$ layers.
\item[{$\ddagger$}] The number of times the energy calculation was repeated in the quantum experiments and noisy simulations, respectively.
\end{tablenotes}
 \label{shots_tabel}
\end{threeparttable}
\end{table}

\section{Results and discussion} \label{results}
In Figs.~\ref{fig:lih_og_h2}-\ref{fig:benzene}, we show the performance of tiled M0 for ground state energy calculations with the tUPS Ansatz on the aforementioned molecular systems. For a given molecule, we use the terms "experiment" and "energy calculation" interchangeably to refer to an energy calculation that used the number of energy measurement shots listed in Table~\ref{h_shots_tabel}. We compare our results to exact/noiseless ground state energies, $E_{\mathrm{exact}}$, calculated with an ideal state vector simulator, i.e. noiseless circuit simulations with infinite shots, so we use the word "exact" in this context to mean energies that would be obtained with flawless error mitigation. Energies obtained from probability vectors that have undergone no error mitigation are referred to as raw energies. Energies obtained from tiled M0 and M0 are denoted $\Delta E_{\mathrm{tM0}}=E_{\mathrm{tM0}}-E_{\mathrm{exact}}$ and $\Delta E_{\mathrm{M0}}$, respectively, and we occasionally refer to the energy differences as energy errors.  We shall refer to an experiment that exclusively used results from a quantum computer to calculate the electronic energy as a quantum experiment. An experiment performed with simulated device noise on a classical computer is referred to as a noisy simulation experiment or, in short, a simulation.  All energy calculations were carried out on separate days for the quantum experiments, and new confusion matrices were constructed each time. New matrices were also constructed for each simulation. Results from all individual energy calculations including condition numbers for the matrices from the quantum experiments are shown in Figs.~\ref{fig:lih_all}-\ref{fig:benzene_simulator_experiments} in the SI with bootstraps for some experiments shown in Figs.~\ref{fig:lih_bootstrap}-\ref{fig:but_h2o_benz_bootstrap}. Figs.~\ref{fig:lih_og_h2}-\ref{fig:benzene} show energy averages over all experiments with error bars that represent sample standard deviations. Numerical values for all standard deviations are given in the figure texts of Figs.~\ref{fig:lih_all}-\ref{fig:benzene_simulator_experiments} in the SI along with RMSD values for tiled M0 and M0 experiments. A few results have been excluded from the main text based on excessive noise which is evident from the large condition numbers of the associated confusion matrices (Table~\ref{s1_meget_stoej} in the SI). In the SI, we also show the effect of using only REM in the LiH quantum experiments. The REM energy can always be obtained as a byproduct in tiled M0 since tiled M0 requires $\mathbf{A}^{\text{REM}}$ alongside $\mathbf{A}^{\text{gate}}$. Note, however, that the obtained REM energy will not be exact for systems of more than $4$ qubits since $\mathbf{A}^{\text{REM}}$ is approximated through Eq.~\eqref{adv_rem} in tiled M0 (i.e. the approximated $\mathbf{A}^{\text{REM}}$ does not include full correlation between all qubits). We observe that the effect of using REM alone is minor compared to M0 and tiled M0, so we only show it for LiH in the SI where the energy errors are reduced by less than $10 \, \mathrm{mE_h}$.

\subsection{\ce{LiH} and \ce{H2}}
\begin{figure}
\centering
\begin{subfigure}{1.0\textwidth}
  \centering
  \includegraphics[width=1.0\linewidth]{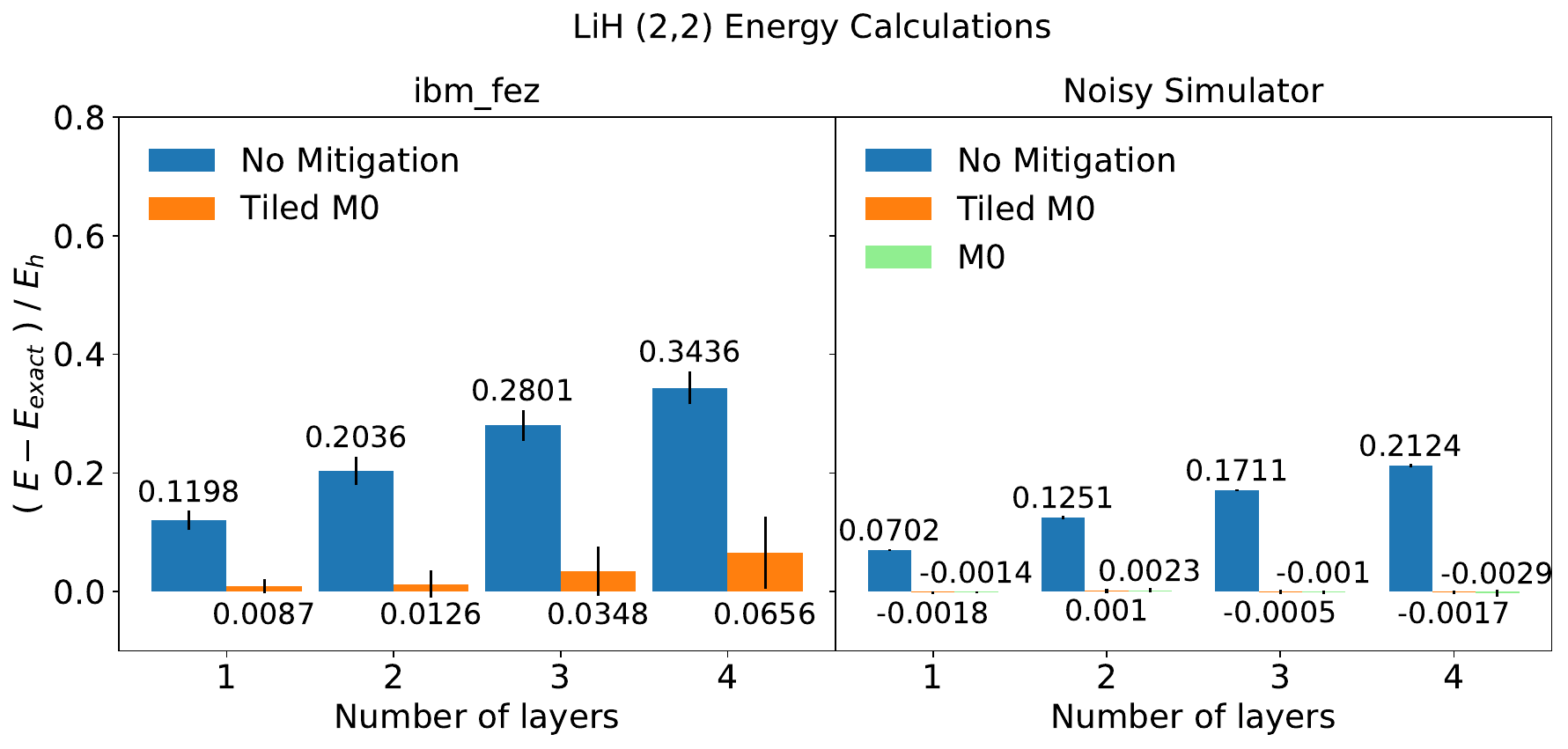}
  \caption{}
  \label{fig:lih}
\end{subfigure}%
\vspace{4mm}
\begin{subfigure}{1.0\textwidth}
  \centering
  \includegraphics[width=1.0\linewidth]{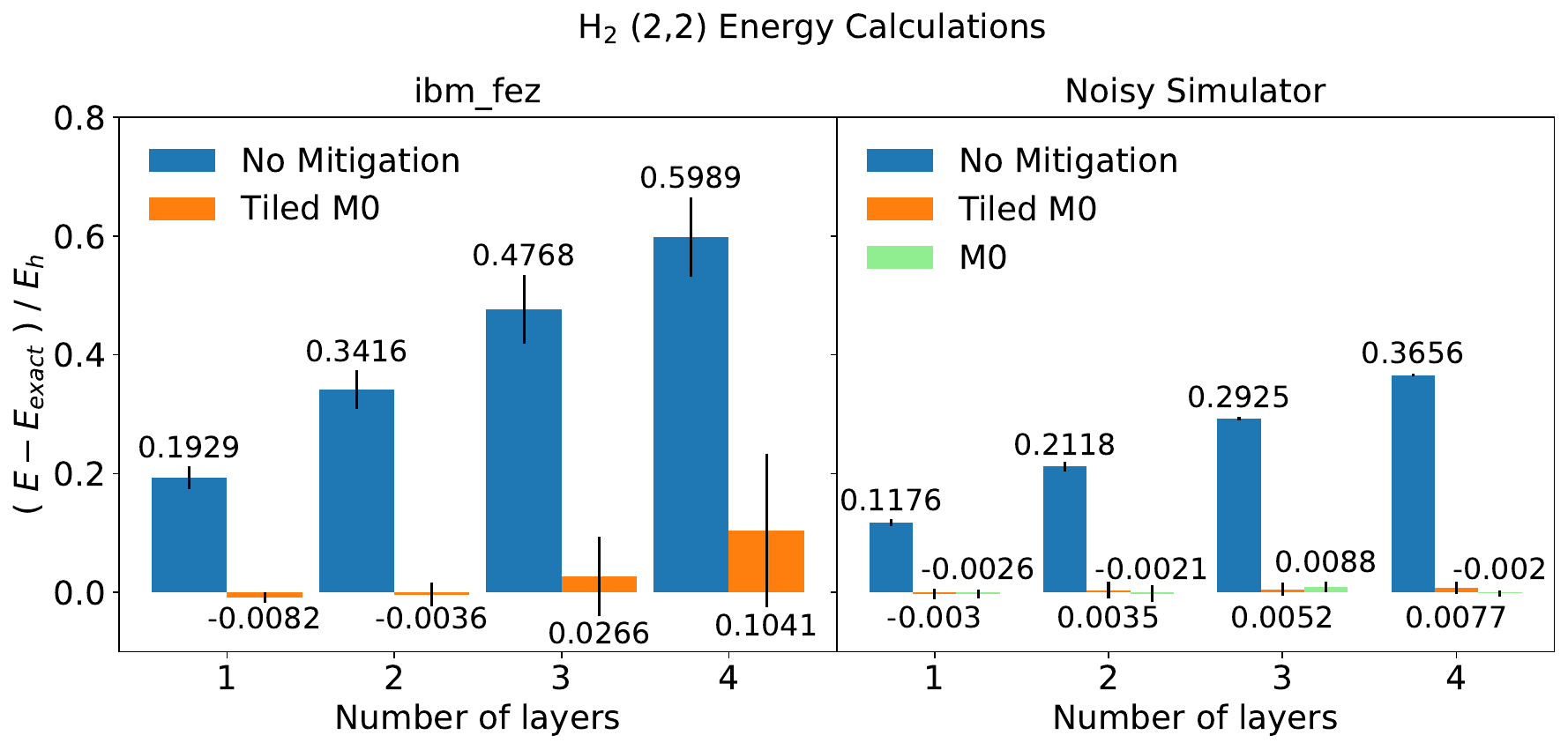}
  \caption{}
  \label{fig:h2}
\end{subfigure}%
\caption{Average results from energy calculations with the tUPS Ansatz on (a) LiH and (b) \ce{H2}. We refer to Table~\ref{shots_tabel} for the number of shots used. Each quantum experiment and simulation was repeated $5$ times. The noise model used for the simulations was imported from \texttt{ibm\_fez}. Note that for both $\ce{LiH}$ and $\ce{H2}$, a single layer is sufficient to reach the ground state energy. For the LiH quantum experiments, the sample standard deviations of $\Delta E_{\mathrm{tM0}}$ are $0.0119$, $0.0227$, $0.0414$ and $0.0608 \, \mathrm{E_h}$ for $1, 2, 3$ and $4$ layers, respectively. The corresponding $\ce{H2}$ standard deviations are $0.0090$, $0.0204$, $0.0668$ and $0.1290 \, \mathrm{E_h}$}
\label{fig:lih_og_h2}
\end{figure}

\begin{figure}
\centering
  \centering
  \includegraphics[width=1\linewidth]{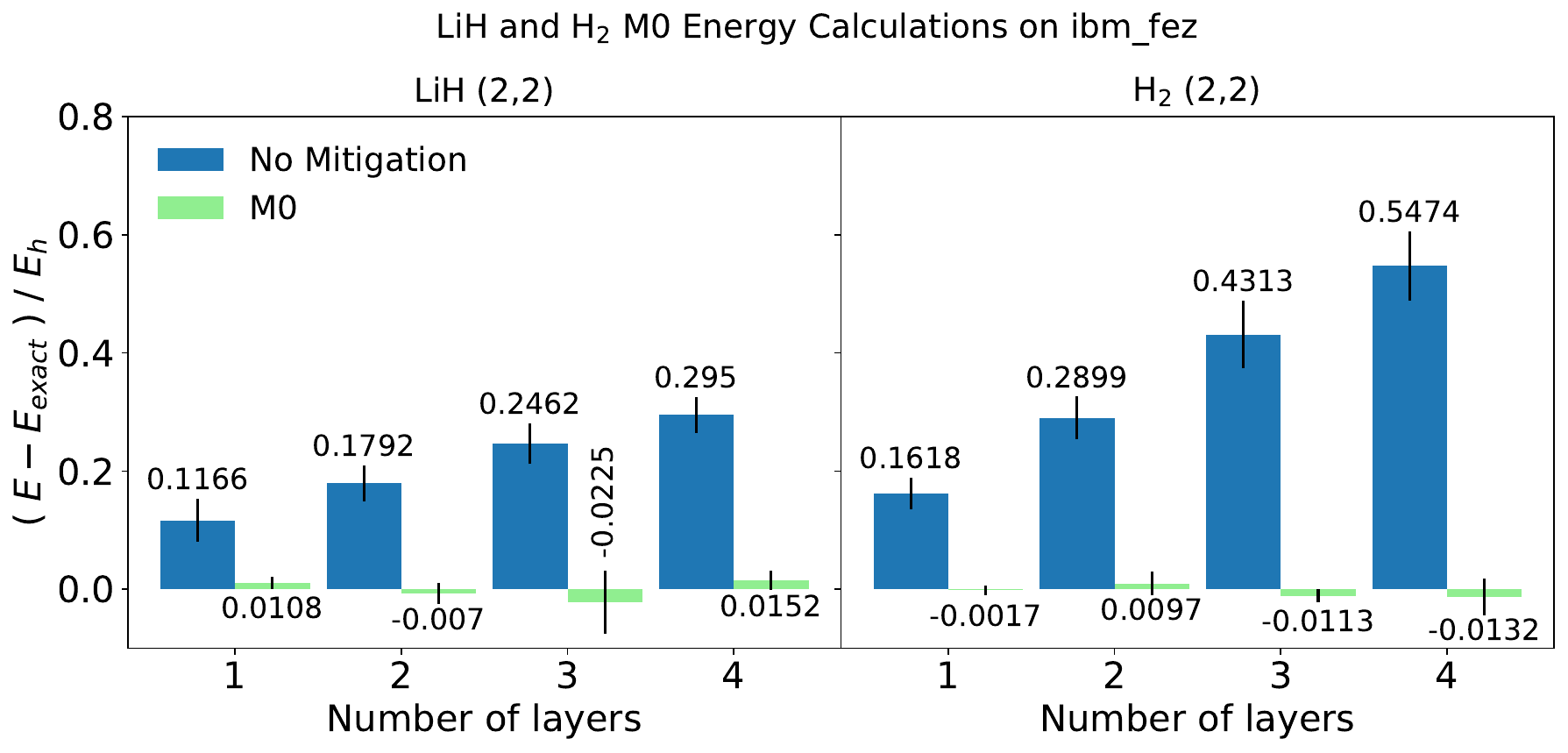}
\caption{Average results from M0 energy calculations with the tUPS Ansatz on LiH and \ce{H2}. We refer to Table~\ref{shots_tabel} for the number of shots used. Each experiment was repeated $5$ times. For LiH, the sample standard deviations of $\Delta E_{\mathrm{M0}}$ are $0.0099$, $0.0182$, $0.0532$ and $0.0161 \, \mathrm{E_h}$ for $1, 2, 3$ and $4$ layers, respectively. The corresponding $\ce{H2}$ standard deviations are $0.0081$, $0.0201$, $0.0109$ and $0.0310 \, \mathrm{E_h}$}
\label{fig:lih_h2_m0}
\end{figure}

Results from experiments on \ce{LiH} and \ce{H2} with $(2,2)$ active spaces are shown in Figs.~\ref{fig:lih_og_h2} and \ref{fig:lih_h2_m0}. The number of CZ gates in the circuits was $39 \times L$, where $L$ is the layer count.

From the \ce{LiH} quantum experiment energies in Fig.~\ref{fig:lih}, we see that tiled M0 is successful in reducing the energy error for all layers compared to the raw results. We get a one order of magnitude reduction in the error at $1$, $2$ and $3$ layers and slightly less at $4$ layers. The best accuracy we achieve with tiled M0 is $\Delta E_{\mathrm{tM0}} = 8.7 \, \mathrm{mE_h}$ at $1$ layer with a sample standard deviation of $s_{\mathrm{tM0}} = 11.9 \, \mathrm{mE_h}$ and $\text{RMSD}_{\text{tM0}} = 13.7 \, \mathrm{mE_h}$. A comparison of the quantum experiments and simulations for LiH shows that we get more accurate results in the simulations for both the raw and tiled M0 energies. Crucially, tiled M0 reduces the energy error between one and two orders of magnitude in the simulations. The best result is for the $3$-layer calculation where the tiled M0 energy deviates only $\Delta E_{\mathrm{tM0}} = -0.5\, \mathrm{mE_h}$ from the exact value with $s_{\mathrm{tM0}} = 3.2 \, \mathrm{mE_h}$ and $\text{RMSD}_{\text{tM0}} = 2.9 \, \mathrm{mE_h}$ . Interestingly, the tiled M0 energies on the simulator do not increase consistently with the number of layers the same way they do for the quantum experiments, evidence of a difference between imported noise models and real quantum hardware noise.

The tests on \ce{H2} in Fig.~\ref{fig:h2} show another example of tiled M0 reliably reducing the energy error for calculations on a $4$ qubit system. The number of CZ gates was the same as for LiH. In both the quantum experiments and simulations, we observe approximately the same improvements as for LiH, getting close to accuracies of $1.6 \, \mathrm{mE_h}$ in the simulations.

In the simulations, tiled M0 performs qualitatively the same as normal M0 (note that tiled M0 and M0 are the same for a $4$-qubit system when the number of layers is $1$; the results differ slightly in our calculations because separate confusion matrices were constructed in the different experiments). The results speaks in favor of the validity of the tiled M0 approximations - particularly the layer and separability approximations - for the imported noise models. In the quantum experiments, the energy errors appear to increase more clearly and rapidly with the number of layers compared to the simulations, although the large standard deviations for the $3$- and $4$-layer quantum experiment energies make reliable interpretation difficult.

To test preliminarily if the observed trend could be a result of a breakdown of the validity of the aforementioned approximations in a real quantum hardware setting, we performed separate M0 quantum experiments on LiH and \ce{H2} which are still feasible for a system of $4$ qubits. In Fig.~\ref{fig:lih_h2_m0}, we show the results. We note that tiled M0 and M0 results are shown on separate figures in the LiH and \ce{H2} quantum experiments because different raw probability vectors were used, unlike the noisy simulations where both tiled M0 and M0 used the same raw probability vectors. The fact that different probability vectors were used in the quantum experiments makes direct comparison between the tiled M0 and M0 results more difficult, since the severity of noise on the \texttt{ibm\_fez} backend that was used could have varied between the experiments (because of noise drift).

Overall, M0 is seen to perform about the same or better than tiled M0 in the quantum experiments. The differences in the $4$-layer calculations are particularly pronounced. For \ce{H2} at $4$ layers, we get $\Delta E_{\mathrm{M0}} = -13.2 \, \mathrm{mE_h}$ ($s_{\mathrm{M0}} = 31.0 \, \mathrm{mE_h}$, $\text{RMSD}_{\text{M0}} = 30.7 \, \mathrm{mE_h}$) compared to $\Delta E_{\mathrm{tM0}} =  104.1 \, \mathrm{mE_h}$ ($s_{\mathrm{tM0}} = 129.0 \, \mathrm{mE_h}$, $\text{RMSD}_{\text{tM0}} = 155.4 \, \mathrm{mE_h}$). Since the $4$-layer calculation is where the layer and separability approximations are tested the most for the LiH and \ce{H2} experiments, and since this is where we see the largest discrepancies between M0 and tiled M0, the results suggest that the feasibility of the tiled M0 approximations for larger layer numbers need further testing on quantum hardware. We note, however, that the average raw energies are lower in the M0 experiments in Fig.~\ref{fig:lih_h2_m0} compared to the tiled M0 experiments in Fig.~\ref{fig:lih_og_h2}. This indicates that the noise was less severe on average in the M0 experiments.
\subsection{Butadiene and \ce{H2O}}
\begin{figure}
\begin{subfigure}{1\textwidth}
  \centering
  \includegraphics[width=1.0\linewidth]{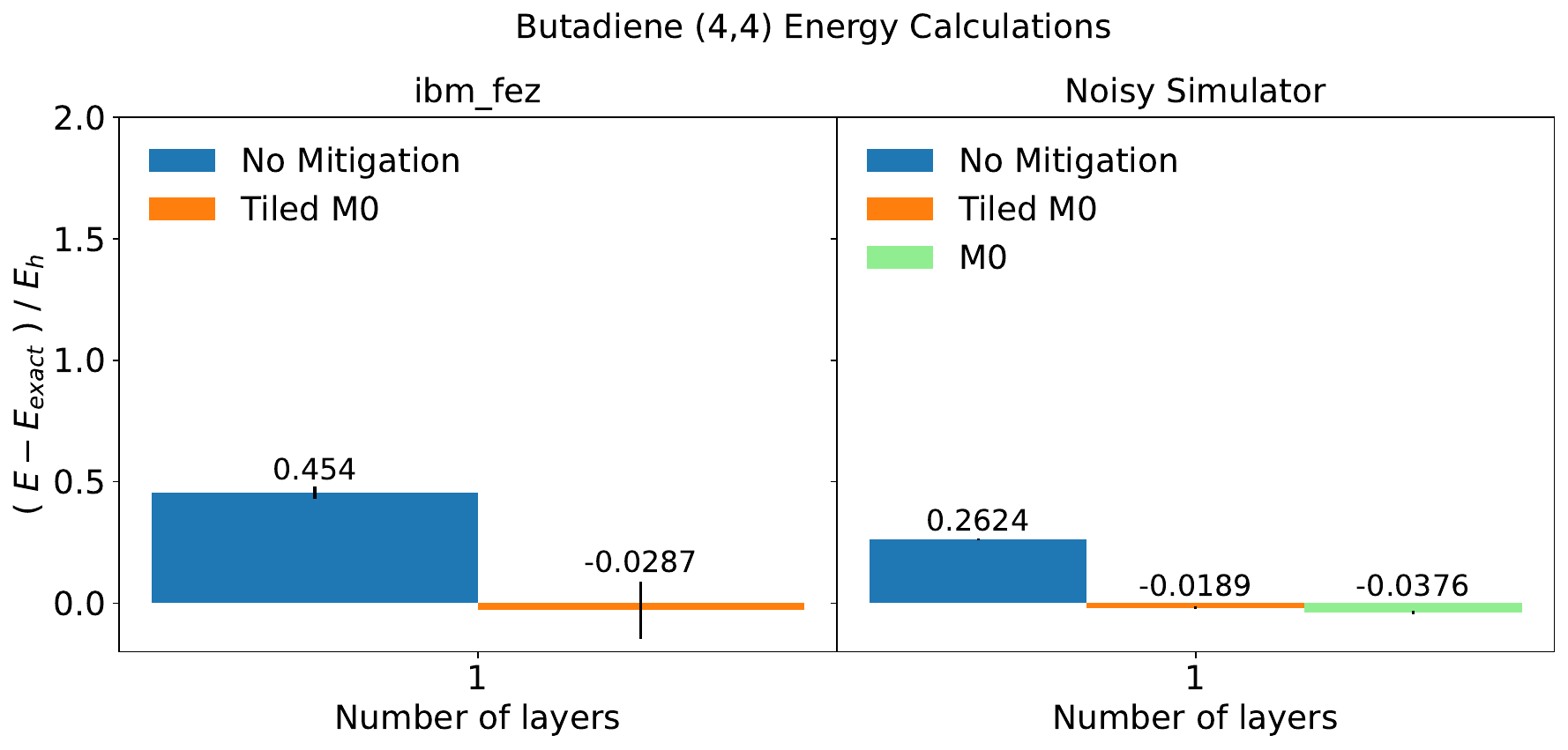}
  \caption{}
  \label{fig:butadiene}
\end{subfigure}%
\vspace{4mm} 
\begin{subfigure}{1\textwidth}
  \centering
  \includegraphics[width=1.0\linewidth]{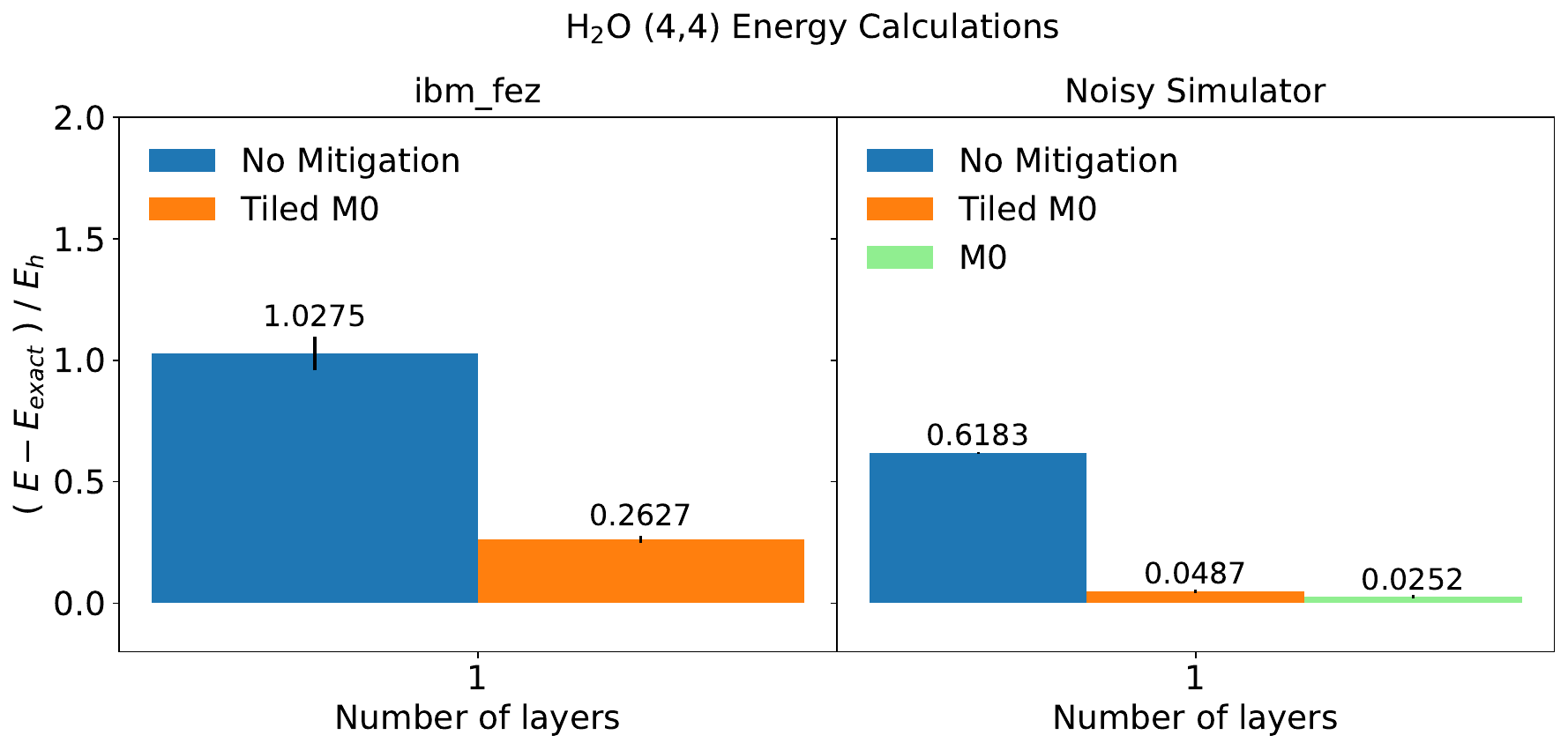}
  \caption{}
  \label{fig:h2o}
\end{subfigure}%
\caption{Average results from energy calculations with the tUPS Ansatz on (a) butadiene and (b) \ce{H2O}. We refer to Table~\ref{shots_tabel} for the number of shots used. Each quantum experiment and simulation was repeated $5$ times for butadiene. For \ce{H2O}, the quantum experiments and simulations were repeated $3$ and $5$ times, respectively. The noise model used for the simulations was imported from \texttt{ibm\_fez}. For the quantum experiments, the sample standard deviations of $\Delta E_{\mathrm{tM0}}$ are $0.1185$ and $0.0160 \, \mathrm{E_h}$ for butadiene and \ce{H2O}, respectively.}
\label{fig:h2o_og_butadiene}
\end{figure}

Fig.~\ref{fig:h2o_og_butadiene} shows results from energy calculations on butadiene and \ce{H2O} with $(4,4)$ active spaces and $151-157$ CZ gates in the $1$-layer tUPS Ansatz (the exact number depends on the qubit mapping). Again, tiled M0 reduces the energy errors by approximately one order of magnitude, e.g.\ for butadiene from $\Delta E_{\mathrm{raw}} = 454.0 \, \mathrm{mE_h}$ ($s_{\mathrm{raw}} = 25.6 \, \mathrm{mE_h}$) to $\Delta E _{\mathrm{tM0}}= -28.7 \, \mathrm{mE_h}$ ($s_{\mathrm{tM0}} = 118.5 \, \mathrm{mE_h}$, $\text{RMSD}_{\text{tM0}} = 109.8 \, \mathrm{mE_h}$) in the quantum experiments, and from $\Delta E_{\mathrm{raw}} = 262.4 \, \mathrm{mE_h}$ ($s_{\mathrm{raw}} = 4.8 \, \mathrm{mE_h}$) to $\Delta E _{\mathrm{tM0}} = -18.9 \, \mathrm{mE_h}$ ($s_{\mathrm{tM0}} = 6.5 \, \mathrm{mE_h}$, $\text{RMSD}_{\text{tM0}} = 19.6 \, \mathrm{mE_h}$) in the simulations. We note the large standard deviation and RMSD associated with the tiled M0 quantum experiment energy and refer to Figs.~\ref{fig:butadiene_experiments} and \ref{fig:but_h2o_benz_bootstrap} in the SI for more details on the individual experiments.

For the tests on water, better energies are obtained with tiled M0 in the simulations, where we see a reduction from $\Delta E_{\mathrm{raw}} = 618.3 \, \mathrm{m E_h}$ ($s_{\mathrm{raw}} = 3.82 \, \mathrm{mE_h}$) to $\Delta E_{\mathrm{tM0}} = 48.7 \, \mathrm{mE_h}$ ($s_{\mathrm{tM0}} = 5.94 \, \mathrm{mE_h}$, $\text{RMSD}_{\text{tM0}} = 49.0 \, \mathrm{mE_h}$) compared to $\Delta E_{\mathrm{raw}} =  1027.5 \, \mathrm{mE_h}$ ($s_{\mathrm{raw}} = 70.0 \, \mathrm{mE_h}$) and $\Delta E_{\mathrm{tM0}} = 263 \, \mathrm{mE_h}$ ($s_{\mathrm{tM0}} = 16.0 \, \mathrm{mE_h}$, $\text{RMSD}_{\text{tM0}} = 263 \, \mathrm{mE_h}$) in the quantum experiments. The quantum experiments only improve the energy error by a factor of $4$. We tentatively suggest that noise drift and noise fluctuations contributed to the worse performance - a point that we discuss further in Sec.~\ref{noise}. Highly non-local noise on the quantum device which the locality approximation in tiled M0 will fail to capture is another potential source of error.
\subsection{Benzene}
\begin{figure}
\centering
  \centering
  \includegraphics[width=1\linewidth]{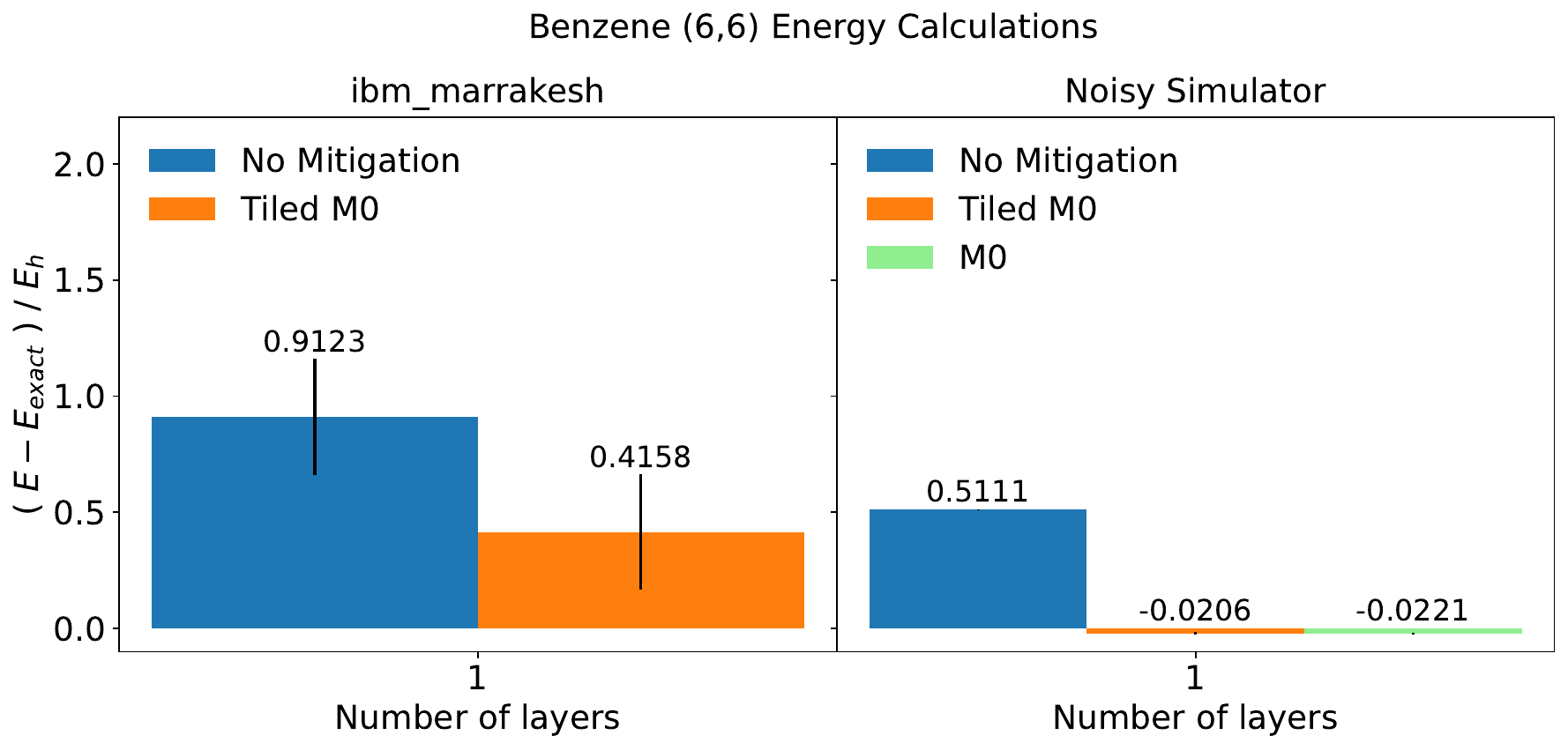}
\caption{Average results from energy calculations with the tUPS Ansatz on benzene. We refer to Table~\ref{shots_tabel} for the number of shots used. Each quantum experiment and simulation was repeated $3$ and $5$ times, respectively. The noise model used for the simulations was imported from \texttt{ibm\_fez}. The sample standard deviation of $\Delta E_{\mathrm{tM0}}$ from the quantum experiments is $0.2492 \, \mathrm{E_h}$.}
\label{fig:benzene}
\end{figure}

Finally, we tested tiled M0 on a large system for current NISQ device capabilities: benzene with a $(6,6)$ active space, corresponding to $12$ qubits, and $287-299$ CZ gates in the $1$-layer tUPS Ansatz. The results are shown in Fig.~\ref{fig:benzene} where we truly see a large difference between the quantum experiments and simulations. In the quantum experiments, tiled M0 reduces the energy error by a factor of $2$ from $\Delta E_{\mathrm{raw}} = 912.3 \, \mathrm{mE_h}$ ($s_{\mathrm{raw}} = 250.0 \, \mathrm{mE_h}$) to $\Delta E_{\mathrm{tM0}} = 415.8 \, \mathrm{mE_h}$ ($s_{\mathrm{tM0}} = 249.2 \, \mathrm{mE_h}$, $\text{RMSD}_{\text{tM0}} = 462.9 \, \mathrm{mE_h}$) while we obtain a better than one order of magnitude improvement on the simulator with both tiled M0 and M0 and better energies overall (e.g. $\Delta E_{\mathrm{tM0}} = -20.6 \, \mathrm{mE_h}$, $s_{\mathrm{tM0}} = 7.1 \, \mathrm{mE_h}$, $\text{RMSD}_{\text{tM0}} = 21.6 \, \mathrm{mE_h}$). We note that the noise levels varied significantly between the different quantum experiments with the minimal and maximal raw energies $671.6 \, \mathrm{mE_h}$ and $1170.8 \, \mathrm{mE_h}$ (Fig.~\ref{fig:benzene_experiments} in the SI). For comparison, the shot noise uncertainty is approximately $5.1 \, \mathrm{mE_h}$ (Table~\ref{shot_noise_tabel}).

While the quantum experiment results are not as impressive (see Sec. \ref{noise} for a discussion), we stress that a quantum experiment study of a $12$-qubit system would be impractical with normal M0, requiring approximately $60$ times as many noise characterization shots for the same confusion matrix accuracy compared to tiled M0.
\subsection{Noise fluctuations} \label{noise}
The differences between the quantum experiment and simulation energies for \ce{H2O} and benzene are significantly larger compared to LiH, \ce{H2}, and butadiene. For both \ce{H2O} and benzene, $\Delta E_{\mathrm{tM0}}$ is one order of magnitude smaller in the simulations.
Bootstraps of individual experiments suggest that the poor results are not due to statistical noise (see Fig.~\ref{fig:but_h2o_benz_bootstrap} in the SI). We suspect that the poor results in the quantum experiments are, in part, due to noise drift and noise fluctuations. The QPU processing times for each \ce{H2O} and benzene experiment were approximately $1$ hour and $15$ minutes and $2$ hours and $5$ minutes, respectively. The noise was characterized at the start of the experiments, and if the noise drifted as we executed the quantum circuits needed to calculate the energies, this would have introduced an error. Since the calculations on LiH, \ce{H2}, and butadiene all took at most $20$ minutes of QPU time, the noise profile was unlikely to change as much over the course of the experiments.

To get an idea of how stable the noise might have been during our experiments, we performed two tests that each lasted $1$ hour and $10$ minutes on \texttt{ibm\_fez}, where we repeatedly executed the $1$-layer tUPS circuit for \ce{H2O} with a $(4,4)$ active space. We ran $150$ jobs in immediate succession with $10^5$ shots each. In Fig.~\ref{drift_stuff} in the supporting information, a plot of a fixed element of the raw probability vector (no error mitigation) from each job result is seen as a function of the time when the job finished executing. A simulation test is also shown for reference. The results from both tests on \texttt{ibm\_fez} indicate that the noise was moderately stable over the $1$ hour and $10$ minute durations, but with a period of about $10$ minutes for each test where severe fluctuations occurred.

We tentatively suggest that fluctuations like those, together with small drifts in the noise, are partially responsible for the poor results obtained for \ce{H2O} and benzene in the quantum experiments. As a small step towards testing this hypothesis, we conducted a single quantum experiment on \ce{H2O} where we partitioned the Hamiltonian into five smaller operators or "batches". We then performed separate energy calculations with new noise characterization and confusion matrix construction for each batch instead of one initial noise characterization for the complete Hamiltonian measurement. This means that we did full tiled M0 noise characterization five times throughout the experiment, increasing the noise characterization cost fivefold from $958,656$ shots in the no-recalibration case to $4,793,280$ shots with recalibration. When the individual batch energies from the recalibration experiment are summed, a better overall energy is obtained ($\Delta E_{\mathrm{tM0}} = 126.8 \, \mathrm{mE_h}$, $\Delta E_{\mathrm{raw}} = 920.1 \, \mathrm{mE_h}$). The result tentatively supports the noise drift hypothesis although the simulation energies are still better. In the SI, more details on the recalibration experiment are given in Fig.~\ref{fig:recalibration} and Table~\ref{batches_tabel} and the surrounding text.

\section{Summary}
\label{sec:conclusion}
We have proposed an error mitigation technique, tiled M0, that characterizes and corrects gate and readout noise for tiled Ans\"atze. Our technique is similar to readout error mitigation, but it differs in two main ways: 1) parts of the Ansatz are included when constructing the confusion matrices, an idea adopted from the recently proposed M0 technique, and 2) special to tiled M0, a locality approximation based on the structure of tiled Ans\"atze is introduced. The latter exponentially reduces the scaling in the QPU cost of the noise characterization compared to fully correlated methods. 


In quantum experiments and noisy simulations, we have demonstrated the ability of tiled M0 to reliably improve ground state energy calculations with the tUPS Ansatz for molecular systems varying in size between $4$ and $12$ qubits with up to $299$ CZ gates in the Ansatz. An order of magnitude reduction in the energy error was seen in most tests, and in multiple simulations we obtained energies accurate to within a few millihartrees of the ideal reference values. In the quantum experiments, tiled M0 performed well for LiH and \ce{H2} and moderately well for butadiene with an order of magnitude reduction in the energy error for all systems, but its impact was limited for \ce{H2O} and benzene, only reducing the error by a factor of $2-4$. Based on preliminary findings, we suspect that this is in part due to noise drift and noise fluctuations, which are exacerbated by long QPU processing times. These effects were not present in the corresponding noisy simulations where tiled M0 performed consistently well.

Tiled M0 also performed very similar to M0 in the simulations, suggesting that the assumptions of tiled M0 are reasonable for the imported noise models. The verdict is not as clear in the quantum hardware setting. Comparisons with M0 quantum experiments for LiH and \ce{H2} showed better results in the $3$- and $4$-layer calculations for M0, potentially indicating a drawback of the layer approximation. The inability of tiled M0 to capture non-local noise between tiles may also have contributed to the errors seen for \ce{H2O} and benzene in the quantum experiments.

In summary, the proposed error mitigation technique provides a simple and cost-effective way of mitigating errors for tiled Ans\"atze with a noise characterization cost that is exponentially cheaper than fully correlated methods like M0 or REM. We are especially encouraged by the noisy simulation results where orders of magnitude improvements over the raw energies were obtained in all our tests. We expect tiled M0 to find use on improved quantum hardware where the noise is lower and less prone to drift, and in near-term quantum computing applications.
However, further validation on quantum hardware is needed to determine the extent to which the assumptions underlying the method hold in practice. In future work, we will investigate this issue in greater detail, including the role of drift and noise fluctuations and possible strategies for mitigating their impact.

\begin{acknowledgement}
We acknowledge the financial support of the Novo Nordisk Foundation for the focused research project \textit{Hybrid Quantum Chemistry on Hybrid Quantum Computers} (HQC)$^2$, grant number NNFSA220080996.
We further thank The National Quantum Algorithm Academy (NQAA) under Danish e-infrastructure Consortium (DeiC) for financial support in form of a PhD stipend to OGLR.
\end{acknowledgement}

\begin{suppinfo}
Quantum circuit implementation of a tile in tUPS in terms of common single- and two-qubit gates, exact electronic energies of the studied molecular systems, results from all individual energy calculation experiments, RMSD values for all repeated tiled M0 and M0 energy calculation experiments, bootstraps of energy calculation experiments before and after error mitigation, results from the noise stability tests, results from the recalibration experiment, information on the operator batches in the recalibration experiment, results from calculations with severe noise.
\end{suppinfo}

\bibliography{main}

\end{document}


\newpage
\section{Quantum Circuit for a Tile in the tUPS Ansatz}
\begin{figure}[h!]
\centering
  \centering
  \includegraphics[width=1.0\linewidth]{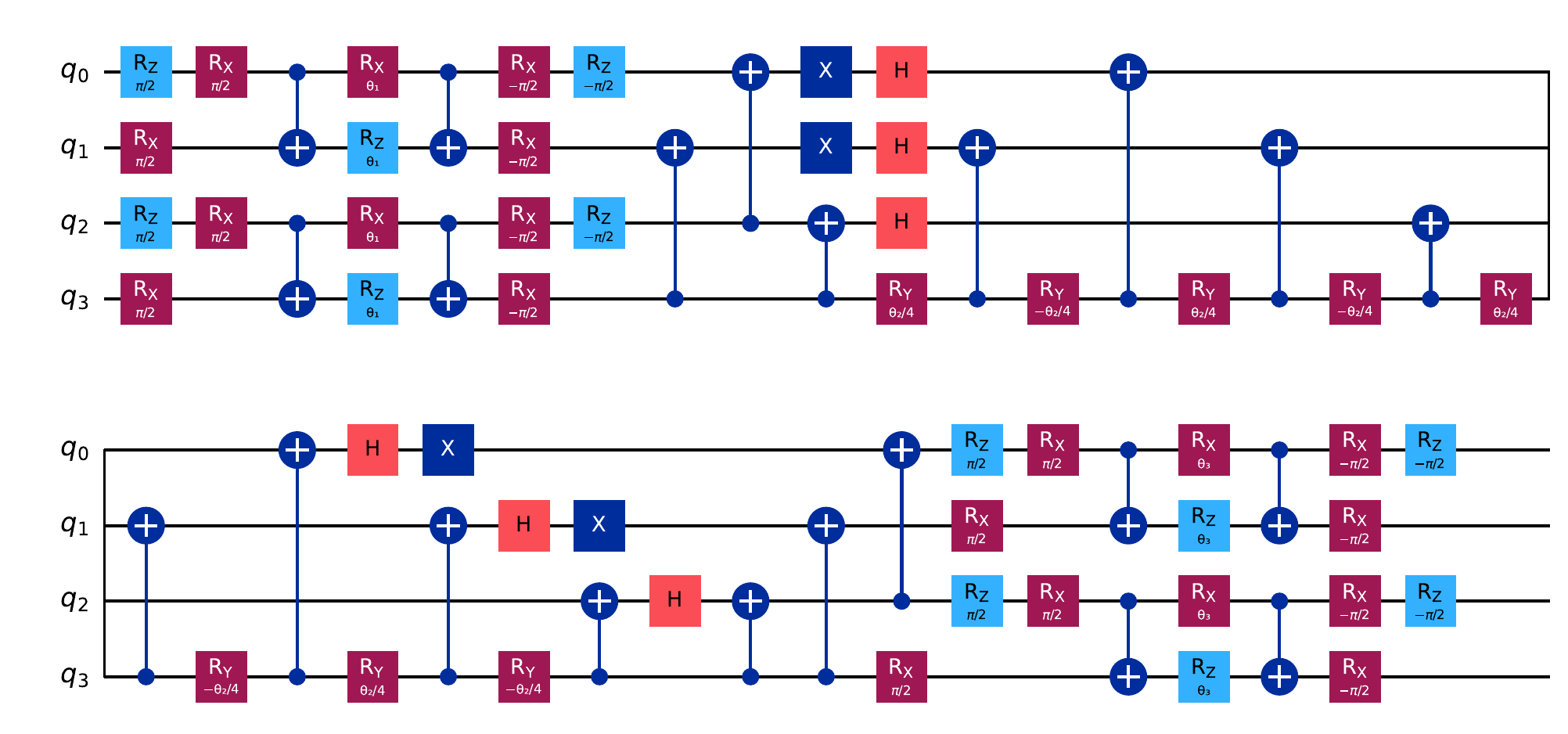}
\caption{The quantum circuit for a tile in the tUPS Ansatz with variational parameters $\theta_1, \theta_2$ and $\theta_3$.}
  \label{fig:tile}
\end{figure}

Fig.~\ref{fig:tile} shows how a single tile in the tUPS Ansatz can be composed in terms of common single and two qubit quantum gates. The variational parameters are $\theta_1, \theta_2$ and $\theta_3$ in the illustration.

\section{Energy experiments} \label{sec:energy_calculations}
\begin{figure}
\centering
  \centering
  \includegraphics[width=1.0\linewidth]{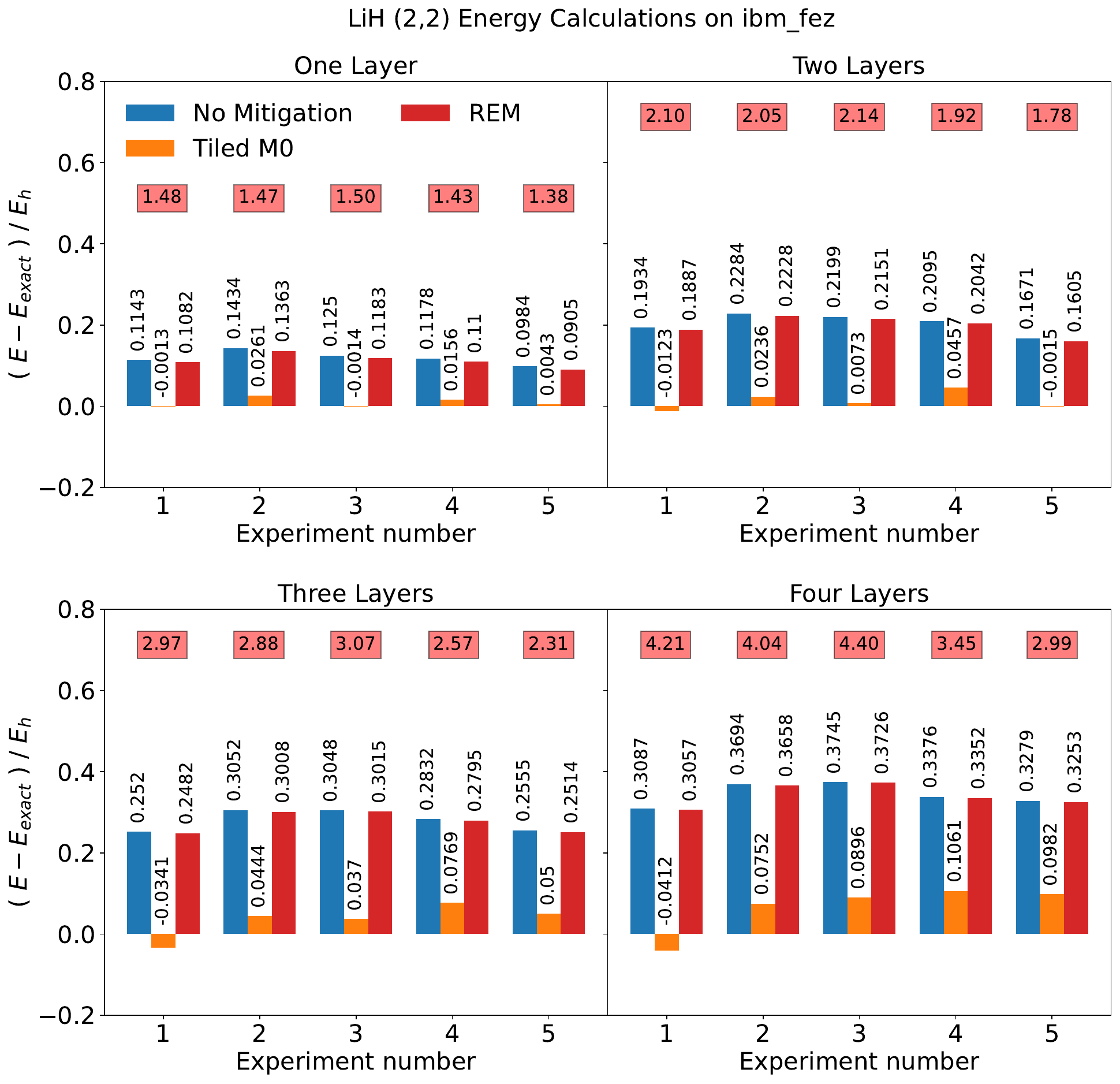}
\caption{Results from LiH (2,2) energy calculation experiments on \texttt{ibm\_fez} at different numbers of layers in the tUPS Ansatz. The condition numbers of the approximated full-system confusion matrices are shown above the energy difference bars for all experiments on the rectangles colored in red. The sample standard deviations of the unmitigated energies are $0.0164$, $0.0242$, $0.0257$ and $0.0279 \, \mathrm{E_h}$ for $1, 2, 3$ and $4$ layers, respectively. The standard deviations of the tiled M0 energies are $0.0119$, $0.0227$, $0.0414$ and $0.0608 \, \mathrm{E_h}$ in the same order of layers. The standard deviations of the REM energies are $0.0166$, $0.0247$, $0.0258$ and $0.0280 \, \mathrm{E_h}$. The RMSD values of the tiled M0 energies are $0.0137$, $0.0239$, $0.0508$ and $0.0852 \, \mathrm{E_h}$.}
  \label{fig:lih_all}
\end{figure}

\begin{figure}
\centering
  \centering
  \includegraphics[width=1.0\linewidth]{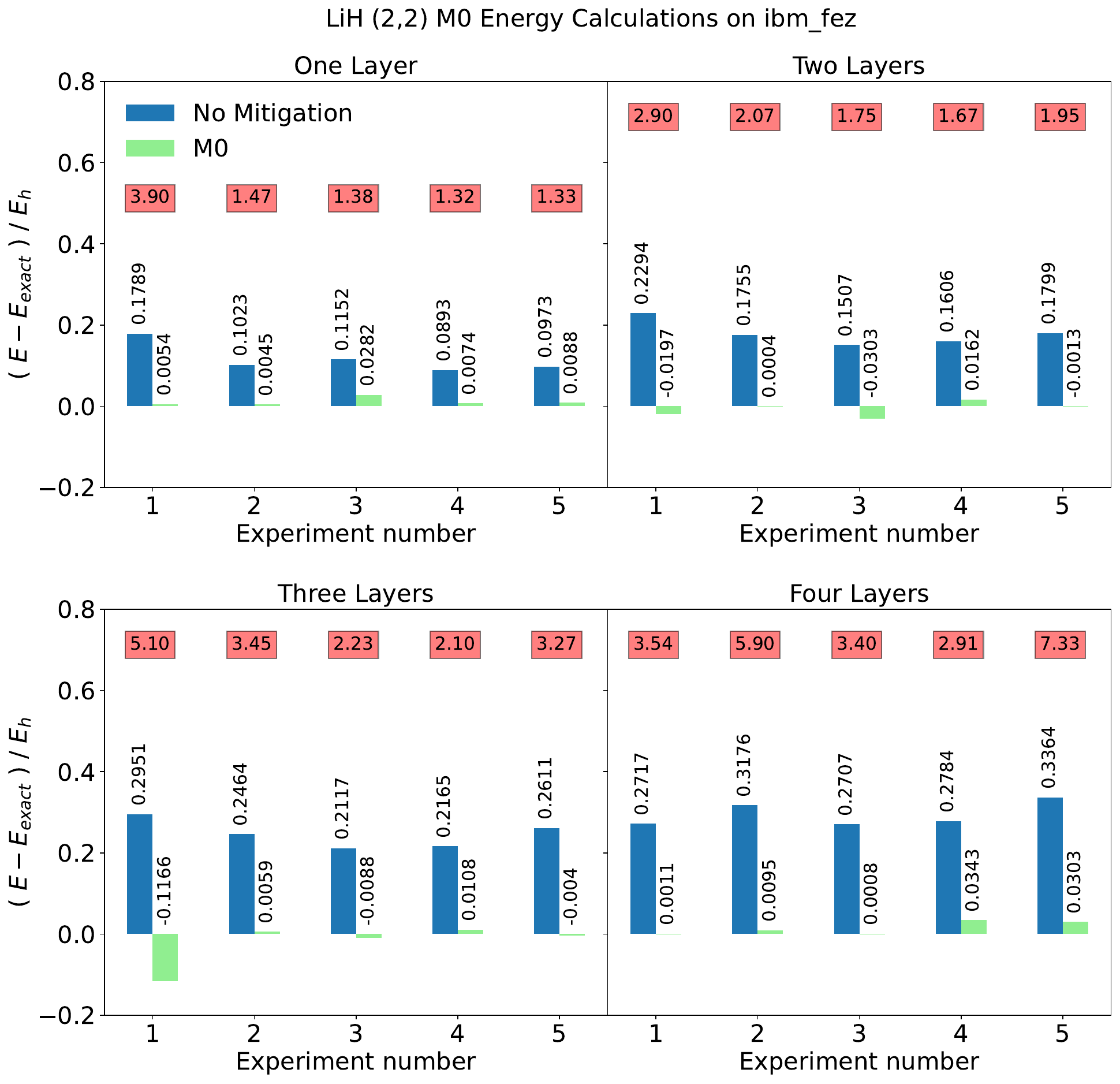}
\caption{Results from LiH (2,2) M0 energy calculation experiments on \texttt{ibm\_fez} at different numbers of layers in the tUPS Ansatz. Condition numbers are shown on the rectangles colored in red. The sample standard deviations of the unmitigated energies are $0.0361$, $0.0304$, $0.0342$ and $0.0302 \, \mathrm{E_h}$ for $1, 2, 3$ and $4$ layers, respectively. The standard deviations of the M0 energies are $0.0099$, $0.0182$, $0.0532$ and $0.0161 \, \mathrm{E_h}$ in the same order of layers. The RMSD values of the M0 energies are $0.0140$, $0.0177$, $0.0526$ and $0.0209 \, \mathrm{E_h}$}
  \label{fig:lih_m0_all}
\end{figure}

\begin{figure}
\centering
  \centering
  \includegraphics[width=1.0\linewidth]{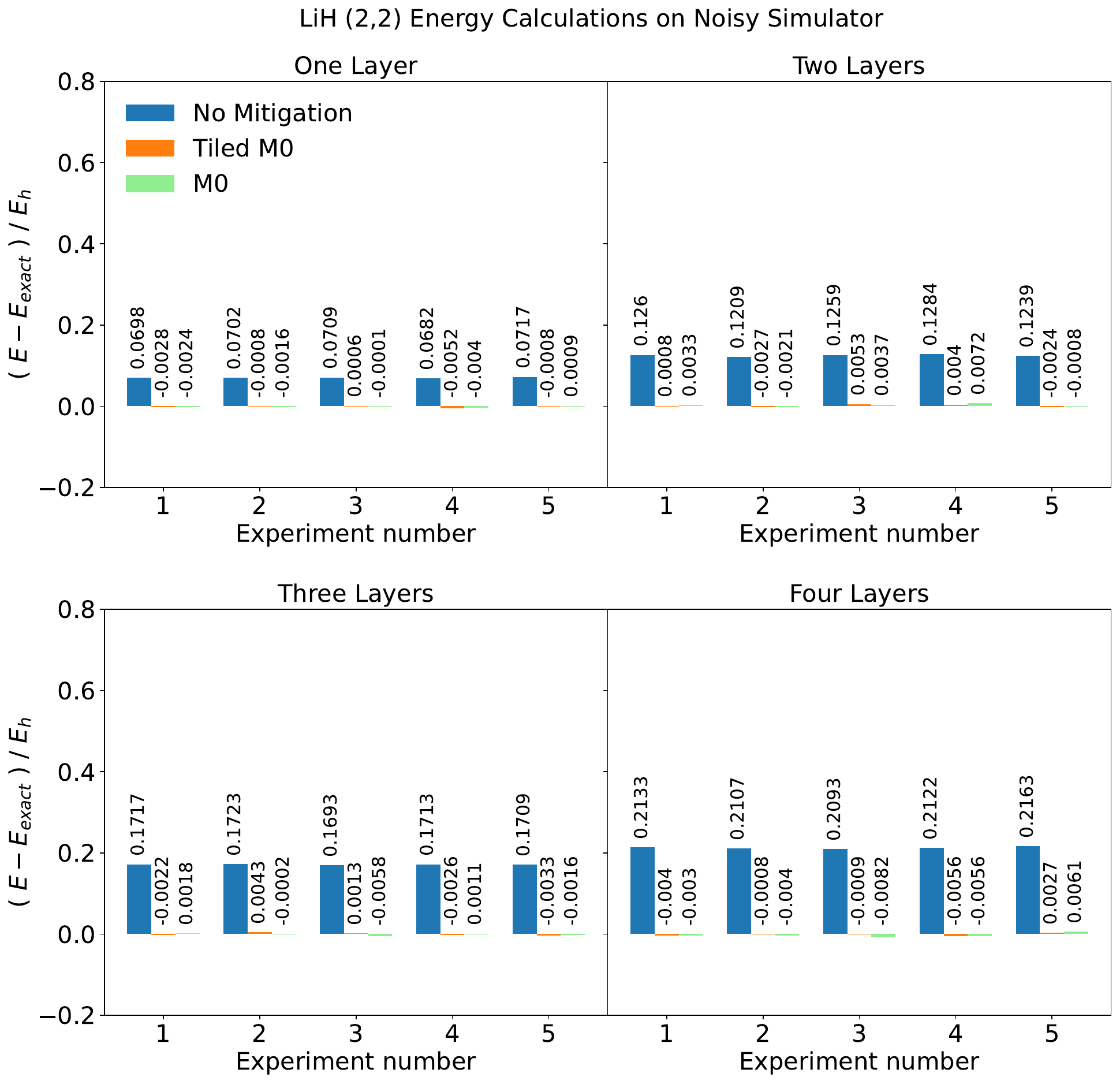}
\caption{Results from LiH (2,2) energy calculation experiments on a noisy simulator at different numbers of layers in the tUPS Ansatz. The noise model was imported from the \texttt{ibm\_fez} backend. The sample standard deviations of the unmitigated energies are $0.0013$, $0.0028$, $0.0011$, and $0.0027 \, \mathrm{E_h}$ for $1, 2, 3$ and $4$ layers, respectively, the standard deviations of the tiled M0 energies are $0.0022$, $0.0037$, $0.0032$, and $0.0032 \, \mathrm{E_h}$, and the standard deviations of the M0 energies are $0.0019$, $0.0037$, $0.0030$, and $0.0054 \, \mathrm{E_h}$ in the same order of layers. The RMSD values of the tiled M0 energies are $0.0027$, $0.0034$, $0.0029$ and $0.0034 \, \mathrm{E_h}$, and the RMSD values of the M0 energies are $0.0022$, $0.0040$, $0.0029$ and $0.0056 \, \mathrm{E_h}$.}
 \label{fig:lih_simulator_all}
\end{figure}

\begin{figure}
\centering
  \centering
  \includegraphics[width=1.0\linewidth]{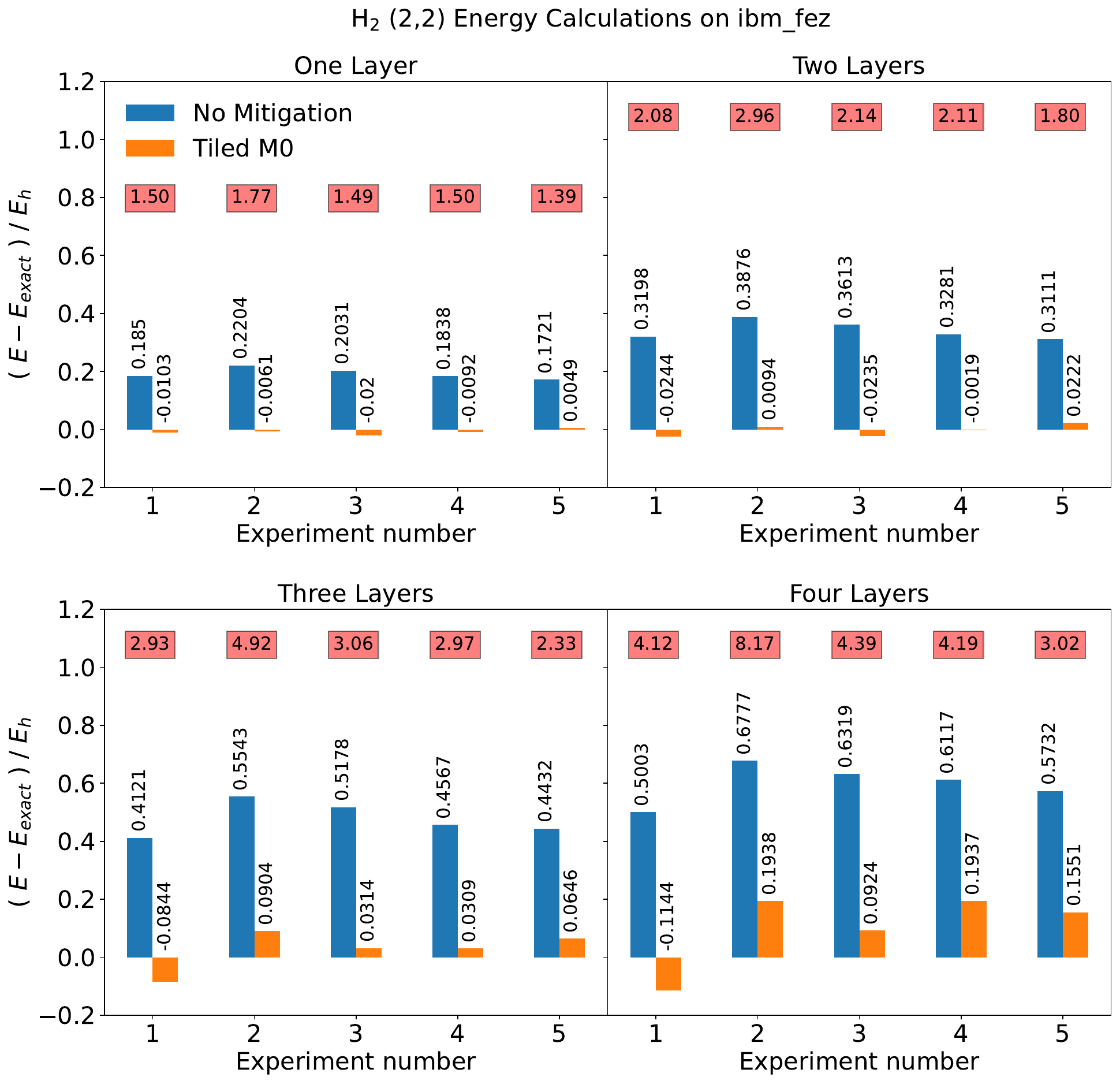}
\caption{Results from $\mathrm{H}_2$ (2,2) energy calculation experiments on \texttt{ibm\_fez} at different numbers of layers in the tUPS Ansatz. Condition numbers are shown on the rectangles colored in red. The sample standard deviations of the unmitigated energies are $0.0190$, $0.0320$, $0.0579$ and $0.0668 \, \mathrm{E_h}$ for $1, 2, 3$ and $4$ layers, respectively. The standard deviations of the tiled M0 energies are $0.0090$, $0.0204$, $0.0668$ and $0.1290 \, \mathrm{E_h}$ in the same order of layers. The RMSD values of the tiled M0 energies are $0.0114$, $0.0186$, $0.0654$ and $0.1554 \, \mathrm{E_h}$.}
  \label{fig:h2_all}
\end{figure}

\begin{figure}
\centering
  \centering
  \includegraphics[width=1.0\linewidth]{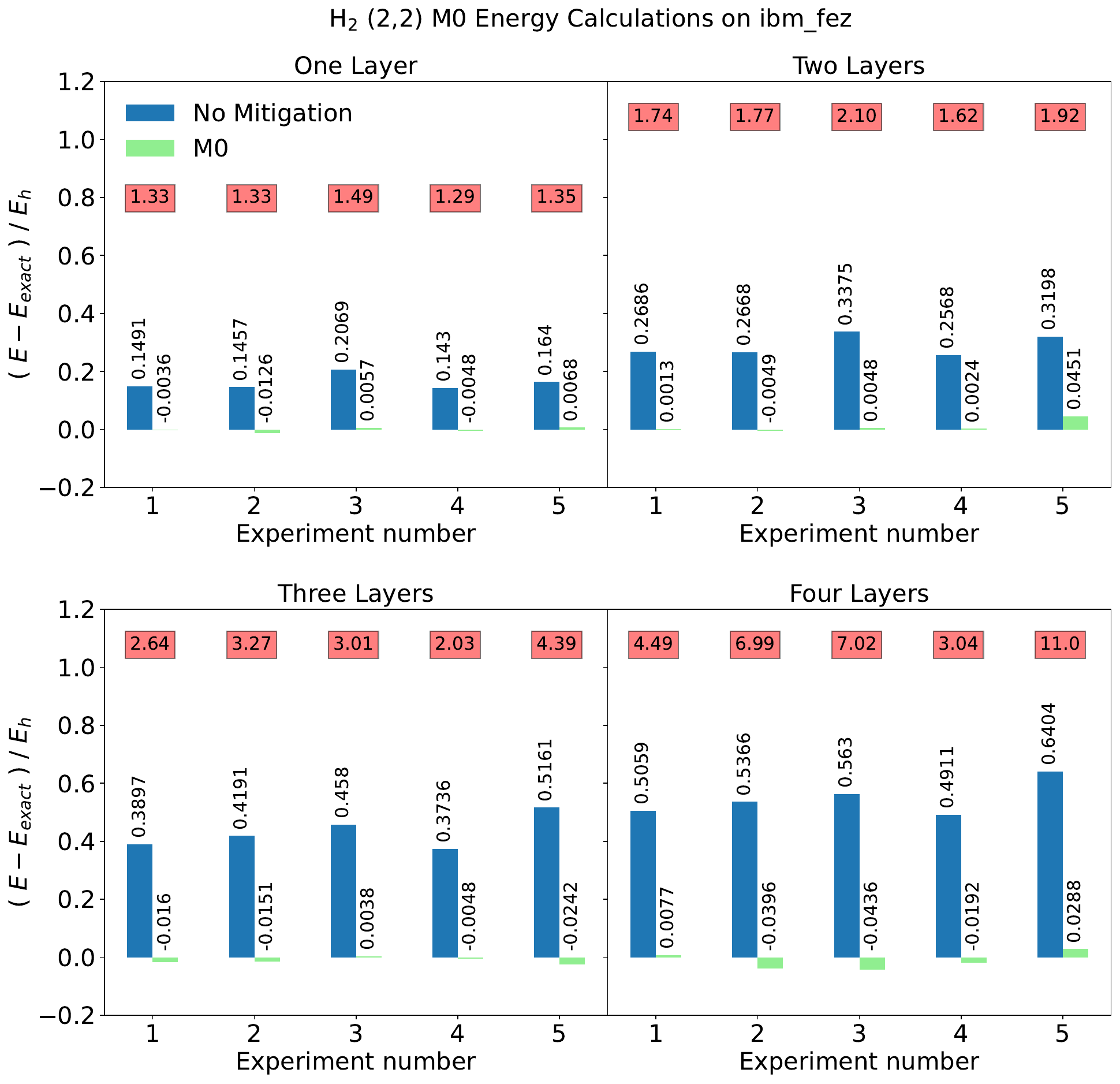}
\caption{Results from $\mathrm{H}_2$ (2,2) M0 energy calculation experiments on \texttt{ibm\_fez} at different numbers of layers in the tUPS Ansatz. Condition numbers are shown on the rectangles colored in red. The sample standard deviations of the unmitigated energies are $0.0265$, $0.0362$, $0.0572$ and $0.0590 \, \mathrm{E_h}$ for $1, 2, 3$ and $4$ layers, respectively. The standard deviations of the M0 energies are $0.0081$, $0.0201$, $0.0109$ and $0.0310 \, \mathrm{E_h}$ in the same order of layers. The RMSD values of the M0 energies are $0.0074$, $0.0205$, $0.0149$ and $0.0307 \, \mathrm{E_h}$.}
  \label{fig:h2_m0_all}
\end{figure}

\begin{figure}
\centering
  \centering
  \includegraphics[width=1.0\linewidth]{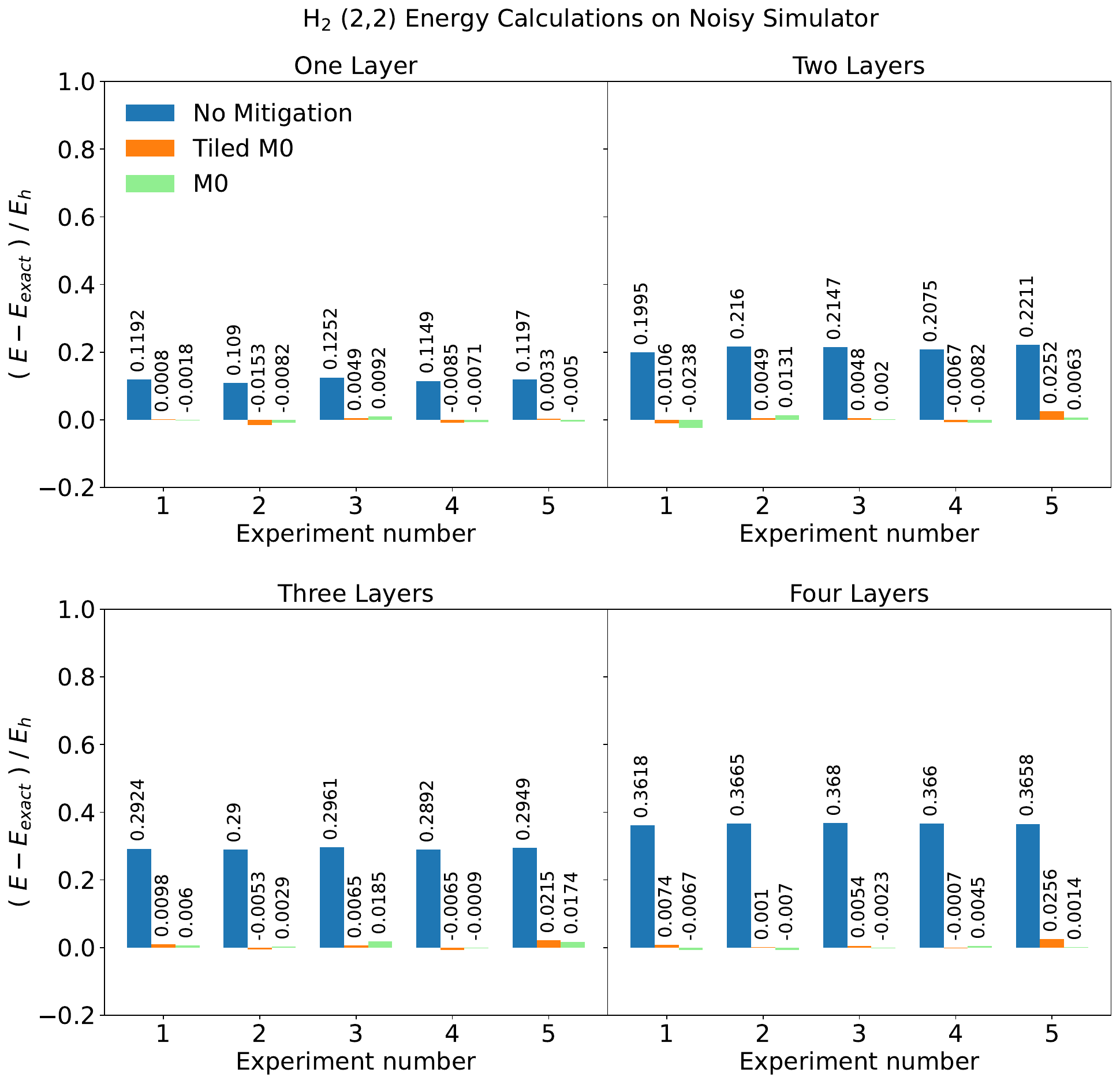}
\caption{Results from $\mathrm{H}_2$ (2,2) energy calculation experiments on a noisy simulator at different numbers of layers in the tUPS Ansatz. The noise model was imported from the \texttt{ibm\_fez} backend. The sample standard deviations of the unmitigated energies are $0.0060$, $0.0084$, $0.0030$, and $0.0023 \, \mathrm{E_h}$ for $1, 2, 3$ and $4$ layers, respectively, the standard deviations of the tiled M0 energies are $0.0087$, $0.0140$, $0.0116$, and $0.0105 \, \mathrm{E_h}$, and the standard deviations of the M0 energies are $0.0070$, $0.01438$, $0.0087$, and $0.0050 \, \mathrm{E_h}$ in the same order of layers. The RMSD values of the tiled M0 energies are $0.0083$, $0.0130$, $0.0116$ and $0.0122 \, \mathrm{E_h}$. The RMSD values of the M0 energies are $0.0068$, $0.0130$, $0.0117$ and $0.0049 \, \mathrm{E_h}$.}
  \label{fig:h2_simulator_all}
\end{figure}

\begin{figure}
\centering
  \centering
  \includegraphics[width=0.6\linewidth]{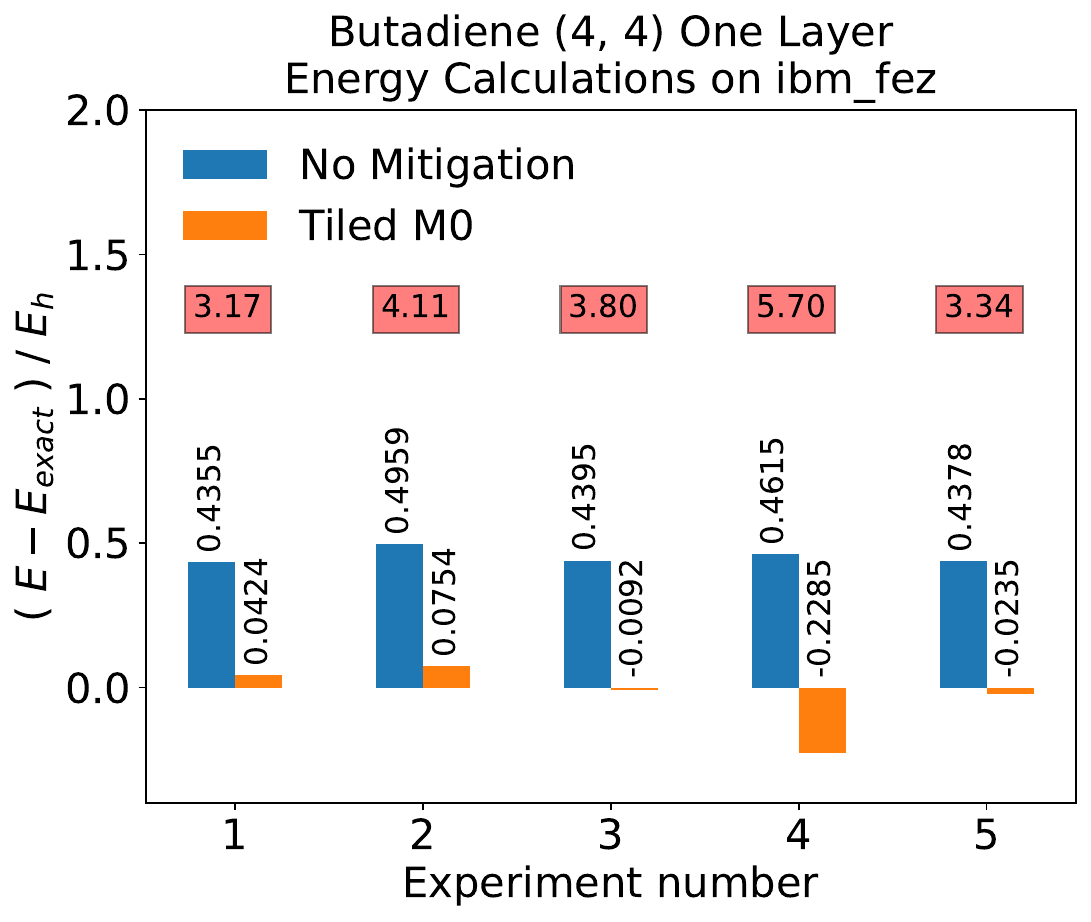}
\caption{Results from butadiene (4,4) energy calculation experiments on \texttt{ibm\_fez} at one layer in the tUPS Ansatz. Condition numbers are shown on the rectangles colored in red. The sample standard deviations of the unmitigated and tiled M0 energies are $0.0256$ and $0.1185 \, \mathrm{E_h}$, respectively. The RMSD of the tiled M0 energies is $0.1098 \, \mathrm{E_h}$.}
  \label{fig:butadiene_experiments}
\end{figure}

\begin{figure}
\centering
  \centering
  \includegraphics[width=0.6\linewidth]{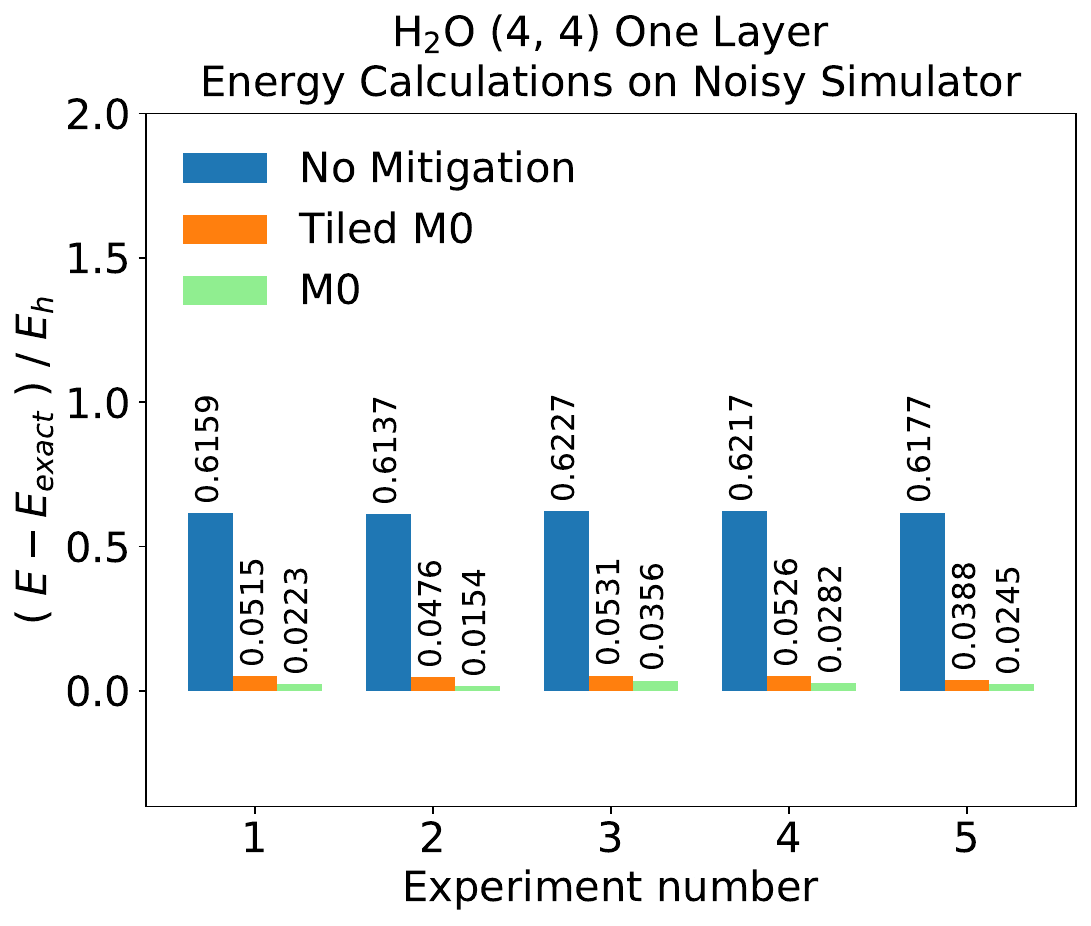}
\caption{Results from $\mathrm{H_2O}$ (4,4) energy calculation experiments on a noisy simulator at one layer in the tUPS Ansatz. The noise model was imported from the \texttt{ibm\_fez} backend. The standard deviations of the unmitigated, tiled M0, and M0 energies are $0.0038$, $0.0059$, and $0.0075 \, \mathrm{E_h}$, respectively. The RMSD values of the tiled M0 and M0 energies are $0.0490$ and $0.0261 \, \mathrm{E_h}$, respectively.}
\label{fig:h2o_simulator_experiments}
\end{figure}

\begin{figure}
\centering
  \centering
  \includegraphics[width=0.6\linewidth]{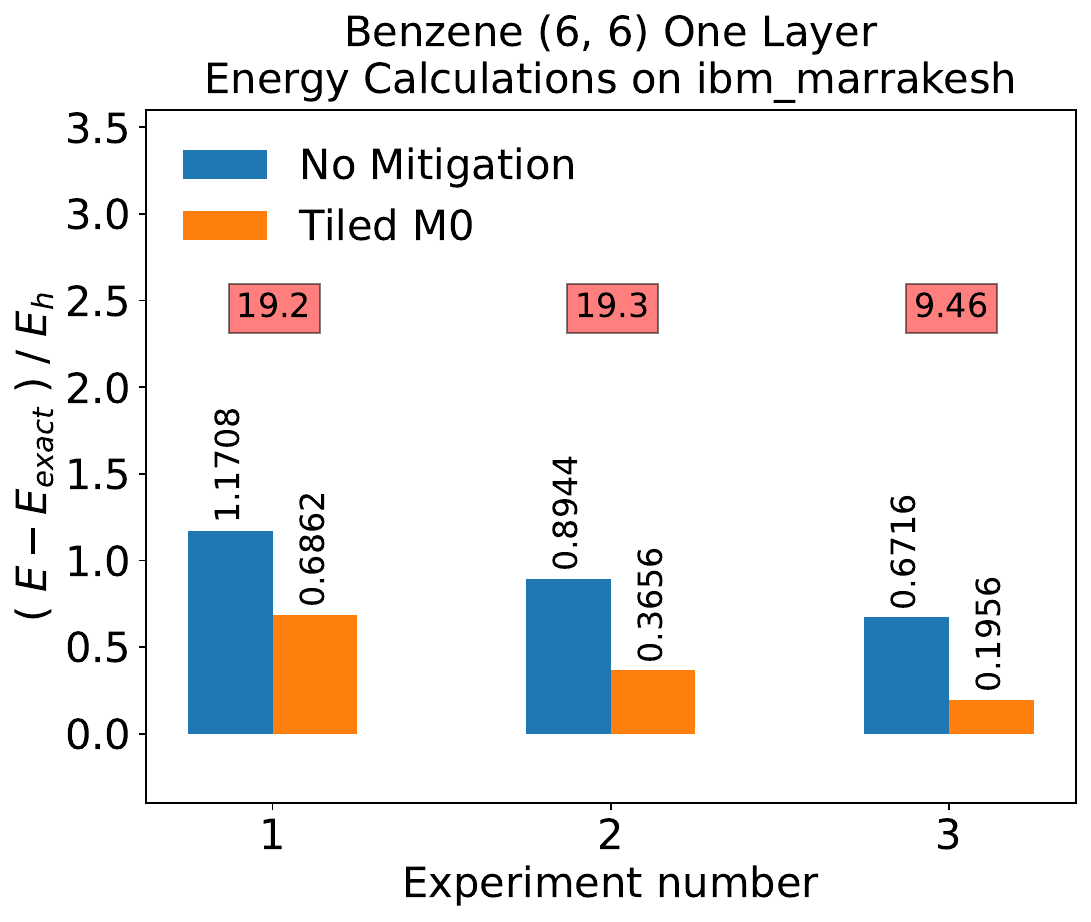}
\caption{Results from benzene (6,6) energy calculation experiments on \texttt{ibm\_marrakesh} at one layer in the tUPS Ansatz. Condition numbers are shown on the rectangles colored in red. The sample standard deviations of the unmitigated and tiled M0 energies are $0.2500$ and $0.2492 \, \mathrm{E_h}$, respectively. The RMSD of the tiled M0 energies is $0.4629 \, \mathrm{E_h}$.}
\label{fig:benzene_experiments}
\end{figure}

\begin{figure}
\centering
  \centering
  \includegraphics[width=0.6\linewidth]{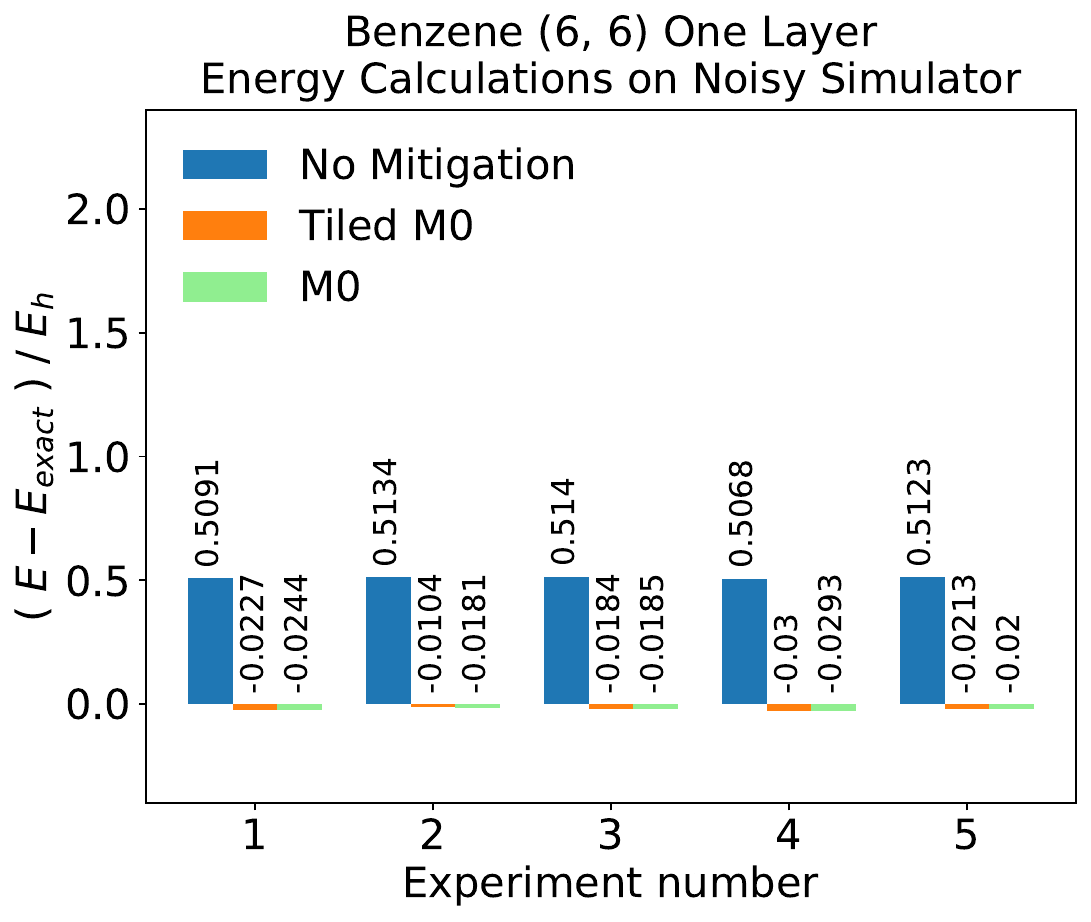}
\caption{Results from benzene (6,6) energy calculation experiments on a noisy simulator at one layer in the tUPS Ansatz. The noise model was imported from the \texttt{ibm\_fez} backend. The \texttt{ibm\_marrakesh} backend was used in the quantum experiments, and preferably the noise model would have been imported from that same backend. The inconsistency is due to a human error, but the differences in the \texttt{ibm\_fez} and \texttt{ibm\_marrakesh} noise models are not expected to be significant. The standard deviations of the unmitigated, tiled M0, and M0 energies are $0.0031$, $0.0071$, and $0.0048 \, \mathrm{E_h}$, respectively. The RMSD values of the tiled M0 and M0 energies are $0.0216$ and $0.0225 \, \mathrm{E_h}$, respectively.}
\label{fig:benzene_simulator_experiments}
\end{figure}

Figs.~\ref{fig:lih_all}-\ref{fig:benzene_simulator_experiments} show results from the individual energy calculation experiments that we performed. The graphs in the main text show averages over all experiments. The total number of shots used in the experiments are given in Table $3$ in the main text. For the LiH quantum experiments, we also show the energies that are obtained using only REM. In Table~\ref{s1_energier}, we show the exact electronic energies that we compare the raw and error mitigated energies with.

The matrix condition numbers of the full-system confusion matrices from the hardware experiments are shown on the red backgrounds on Figs.~\ref{fig:lih_all}-\ref{fig:benzene_simulator_experiments} above the energy difference bars. The full-system matrices for the tiled M0 calculations were approximated according to Eq.~(17) in the main text. Note that for the tiled M0 error mitigation procedure itself, it is not necessary to explicitly construct the approximated full-system matrix. This was done only for the purpose of calculating the condition numbers. The condition numbers for the tile confusion matrices can also be used directly to assess the noise severity.

In the figure texts of Figs.~\ref{fig:lih_all}-\ref{fig:benzene_simulator_experiments}, numerical values for the sample standard deviations of the different energies are given. They were calculated using Bessel's correction. The standard deviations are shown as error bars on the graphs in the main text. Note that noise drift between experiments on quantum hardware may have influenced the standard deviations, adding to the variance already present from shot noise and error mitigation. Bootstraps of individual experiments, from which standard deviations that are free from drift contributions can be estimated, are shown in Sec.~\ref{energy_histograms} below.



\begin{table}[t]
\begin{threeparttable}
\centering\renewcommand\cellalign{lc}
\setcellgapes{3pt}\makegapedcells
\caption{The exact electronic energies in hartree that we compare the raw and error mitigated energies with. By exact we mean with respect to the given level of theory, the number of layers in the tUPS Ansatz (one), the chosen initial state (perfect-pairing), and the specific variational parameters obtained from the classical optimization routine. Note that the energies for LiH and $\ce{H2}$ were the same for all numbers of layers. The energies were calculated with an ideal state vector simulator.}
\centering\renewcommand\cellalign{lcc}
\setcellgapes{3pt}\makegapedcells
\begin{tabularx}{\linewidth}{XXXXX}
\hline
 LiH  &  \ce{H2} & Butadiene & \ce{H2O} & Benzene  \\  \hline 
$-8.87325$ & $-1.85239$ & $-257.17168$ & $-84.00123$ & $-432.43028$ \\ 
\bottomrule
\end{tabularx}
 \label{s1_energier}
\end{threeparttable}
\end{table}

\begin{figure}
\centering
  \centering
  \includegraphics[width=0.6\linewidth]{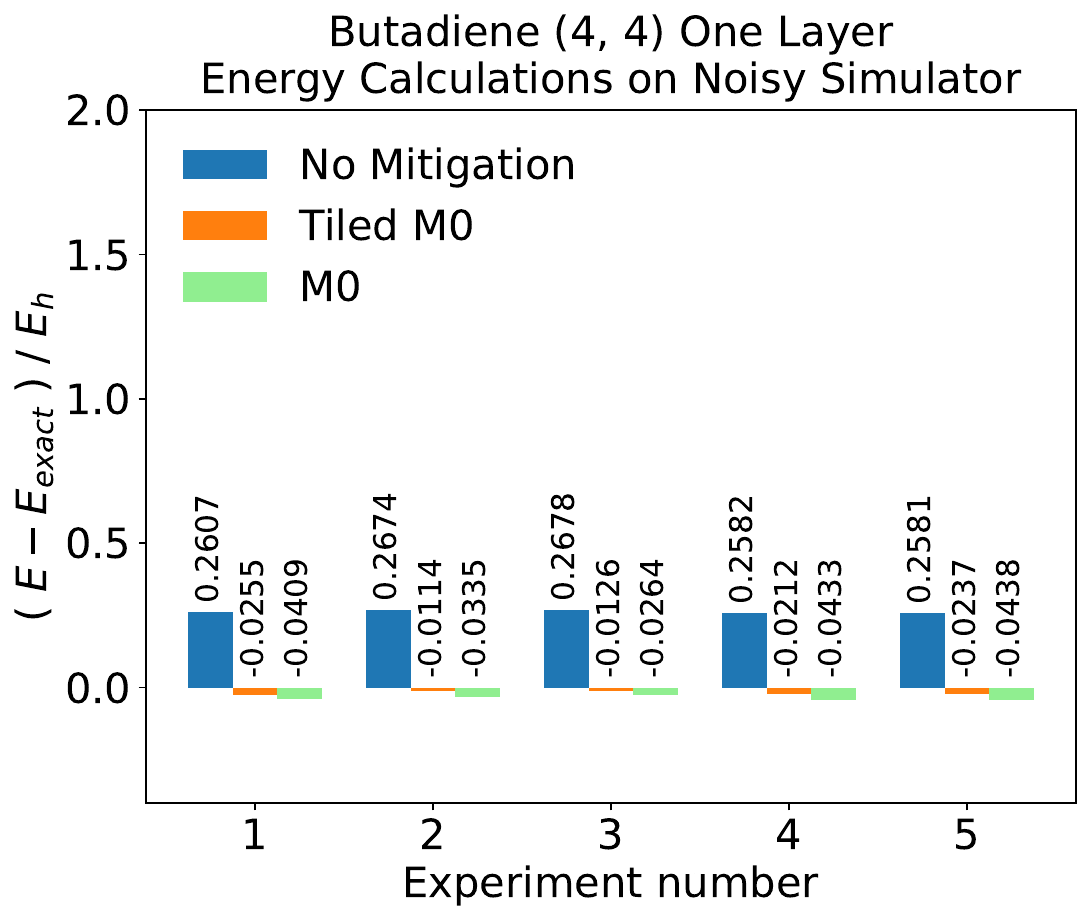}
  \label{fig:butadiene_simulator_experiments}
\caption{Results from butadiene (4,4) energy calculation experiments on a noisy simulator at one layer in the tUPS Ansatz. The noise model was imported from the \texttt{ibm\_fez} backend. The standard deviations of the unmitigated, tiled M0, and M0 energies are $0.0048$, $0.0065$, and $0.0075 \, \mathrm{E_h}$, respectively. The RMSD values of the tiled M0 and M0 energies are $0.0196$ and $0.0382 \, \mathrm{E_h}$, respectively.}
\label{fig:butadiene_simulator_experiments}
\end{figure}

\begin{figure}
\centering
  \centering
  \includegraphics[width=0.6\linewidth]{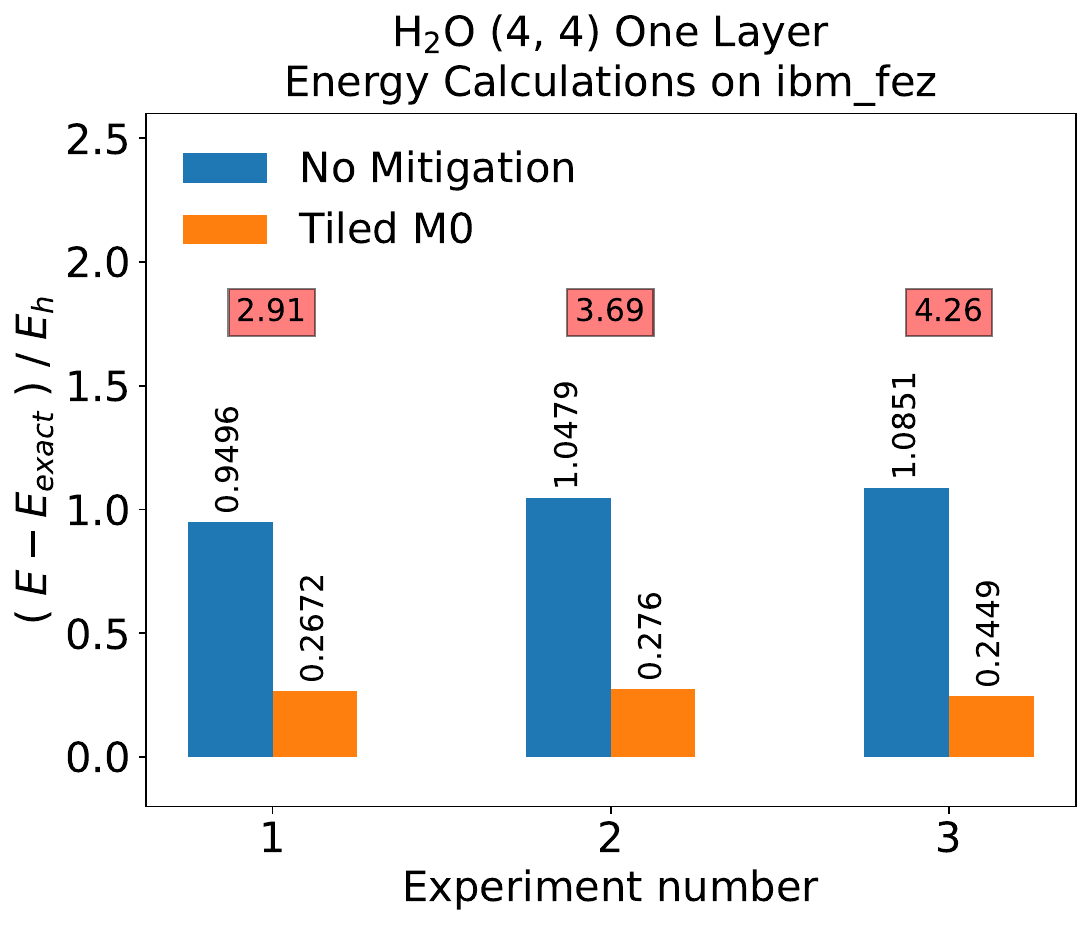}
  \label{fig:h2o_experiments}
\caption{Results from $\mathrm{H_2O}$ (4,4) energy calculation experiments on \texttt{ibm\_fez} at one layer in the tUPS Ansatz. Condition numbers are shown on the rectangles colored in red. The sample standard deviations of the unmitigated and tiled M0 energies are $0.0700$ and $0.0160 \, \mathrm{E_h}$, respectively. The RMSD of the tiled M0 energies is $0.2630 \, \mathrm{E_h}$.}
\label{fig:h2o_experiments}
\end{figure}

\section{Bootstraps} \label{energy_histograms}
\begin{figure}
\centering
  \centering
  \includegraphics[width=1.0\linewidth]{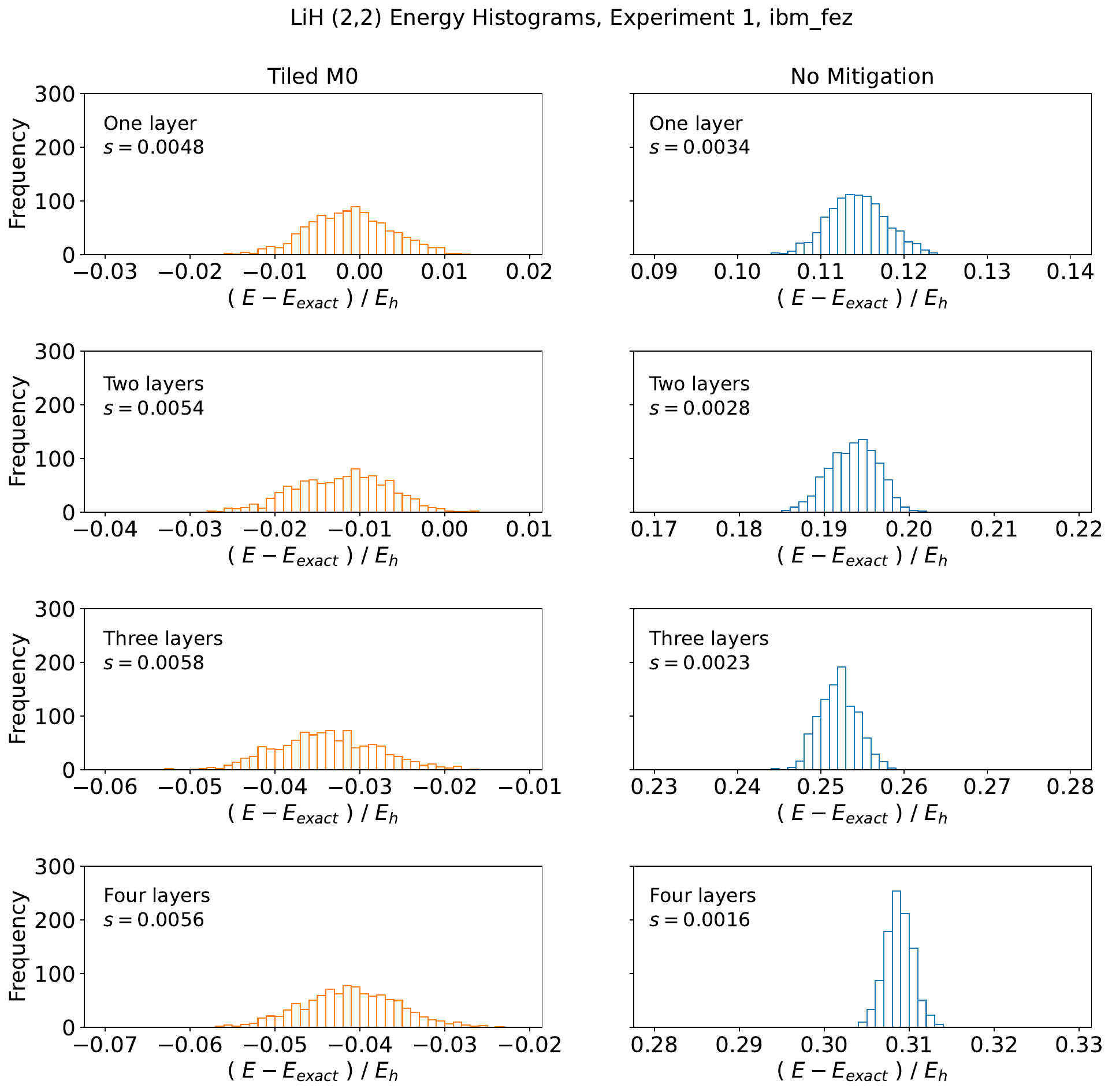}
\caption{Histograms of raw and error mitigated energies calculated via bootstrapping from the LiH experiments labeled $1$ in Fig.~\ref{fig:lih_all}. $10^3$ energy samples were used for each histogram. The total number of shots used for each energy sample is shown in Table 1 in the main text. The bar width is $1 \, \mathrm{mE_h}$. The variable $s$ denotes the sample standard deviation in units of hartree. The ranges on the $x$- and $y$-axes are the same on all histograms.}
  \label{fig:lih_bootstrap}
\end{figure}

\begin{figure}
\centering
  \centering
  \includegraphics[width=1.0\linewidth]{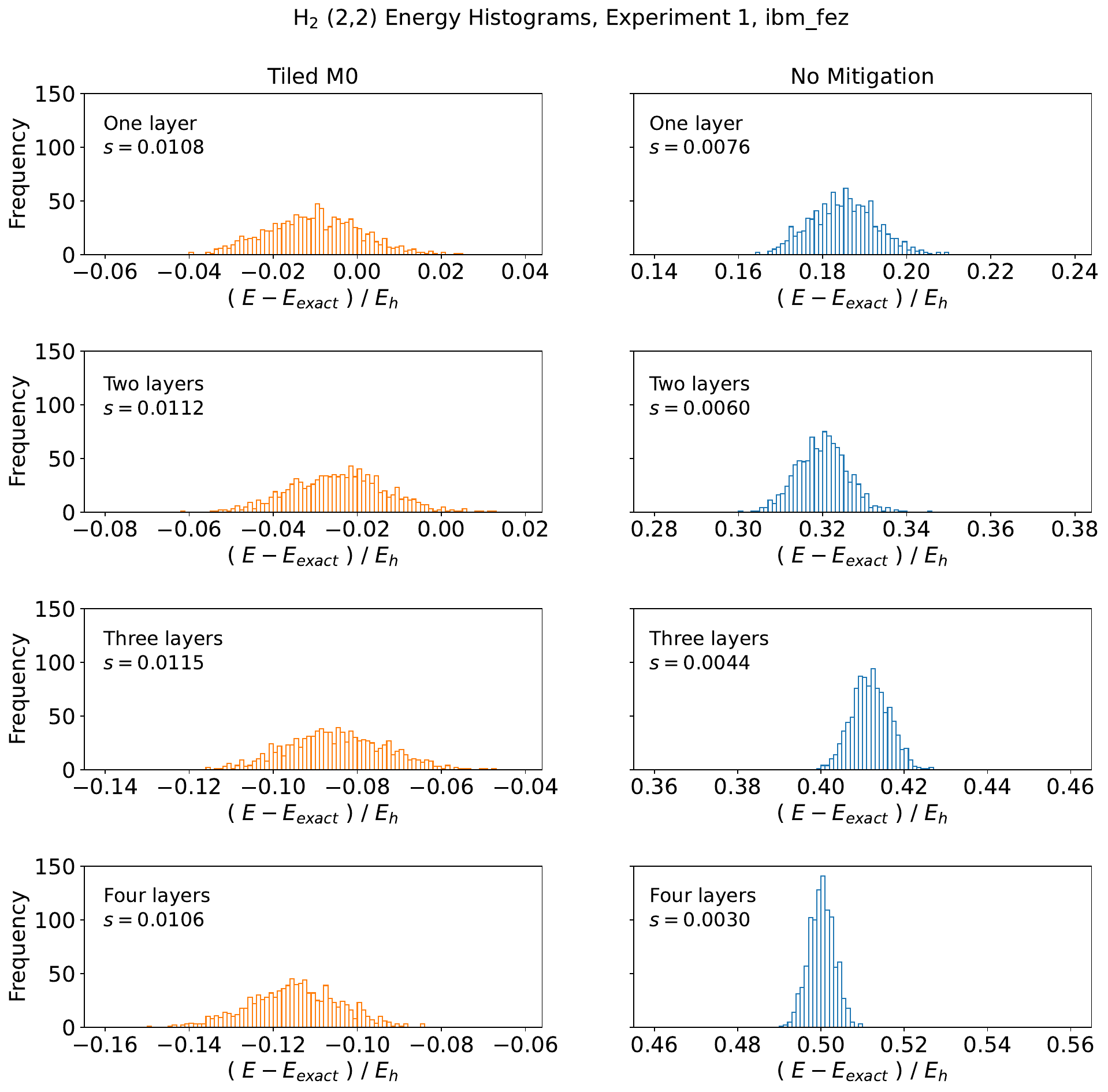}
\caption{Histograms of raw and error mitigated energies calculated via bootstrapping from the $\ce{H2}$ experiments labeled $1$ in Fig.~\ref{fig:h2_all}. $10^3$ energy samples were used for each histogram. The total number of shots used for each layer for each energy calculation is shown in Table 1 in the main text. The bar width is $1 \, \mathrm{mE_h}$. The variable $s$ denotes the sample standard deviation in units of hartree. The ranges on the $x$- and $y$-axes are the same on all histograms.}
  \label{fig:h2_bootstrap}
\end{figure}

\begin{figure}
\centering
  \centering
  \includegraphics[width=1.0\linewidth]{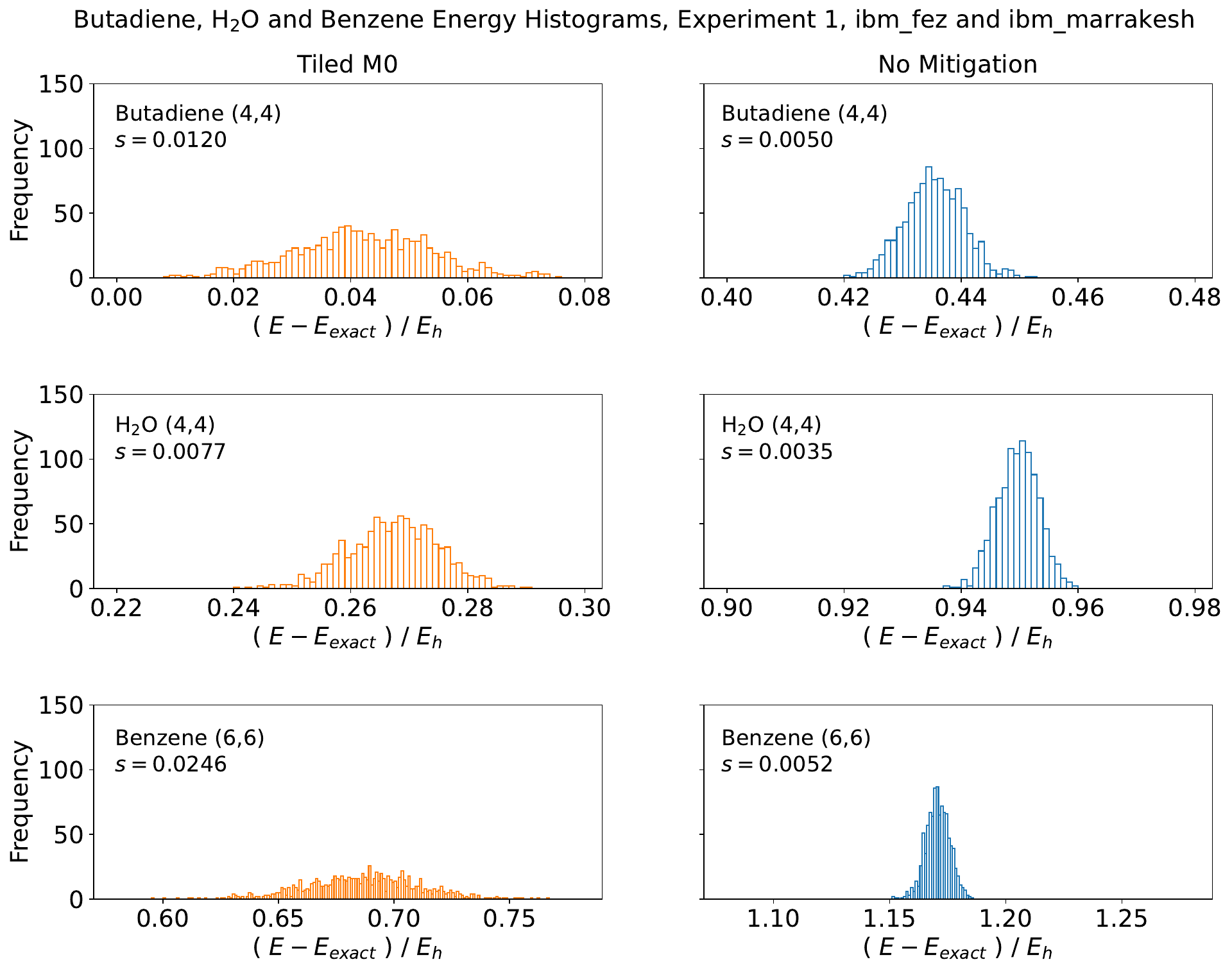}
\caption{Histograms of raw and error mitigated energies calculated via bootstrapping from the butadiene, $\ce{H2O}$ and benzene experiments labeled $1$ in Figures \ref{fig:butadiene_experiments}, \ref{fig:h2o_experiments} and \ref{fig:benzene_experiments}. $10^3$ energy samples were used for each histogram. The total number of shots used for each energy calculation is shown in Table 1 in the main text. The bar width is $1 \, \mathrm{mE_h}$. The variable $s$ denotes the sample standard deviation in units of hartree. Note that the ranges on the $x$-axes for the benzene histograms are different from butadiene and \ce{H2O}.}
  \label{fig:but_h2o_benz_bootstrap}
\end{figure}

Figs.~\ref{fig:lih_bootstrap}-\ref{fig:but_h2o_benz_bootstrap} show energy histograms obtained via bootstrapping for some of the quantum experiments in Sec.~\ref{sec:energy_calculations}. From each raw probability vector associated with a certain QWC-group, a number of samples were drawn equal to the number of shots used to construct the probability vector. New probability vectors were constructed from the samples and those were used to compute resampled raw and error mitigated energies. The error mitigated energies were obtained through the standard tiled M0 procedure, and the tile confusion matrices were not resampled. This means that the bootstraps do not include uncertainties that will arise in practice from variability in the confusion matrices. In total, $10^3$ resampled energies were used for each histogram. The total number of shots used for each sampled energy are shown in Table~1 in the main text. The standard deviations calculated from the bootstrap data are shown on the histograms in units of hartree both before and after error mitigation.

\section{Noise drift and noise fluctuations}
\begin{figure}
\centering
  \centering
  \includegraphics[width=0.6\linewidth]{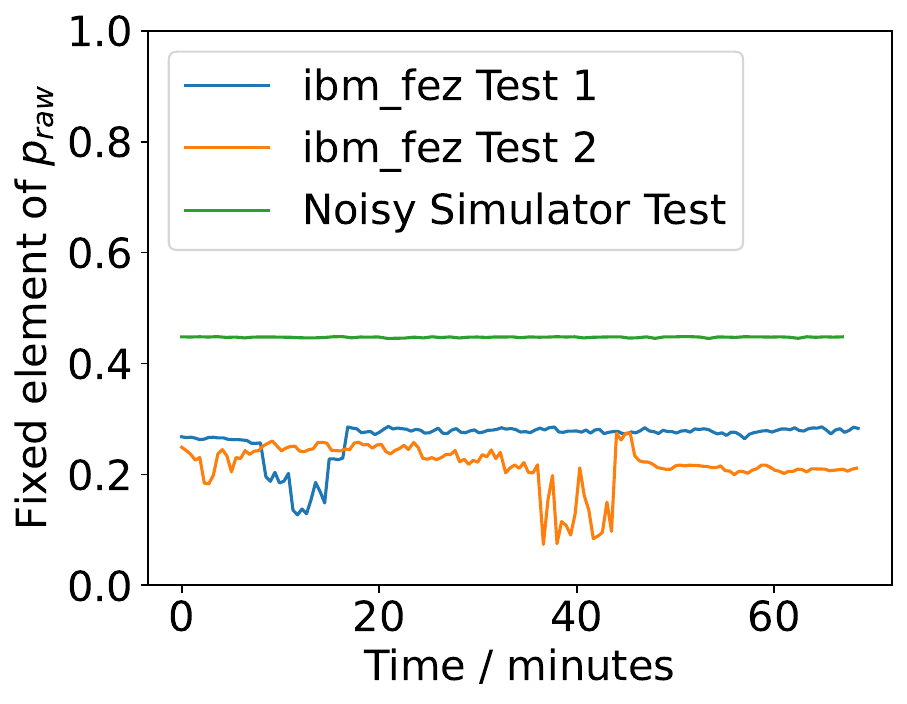}
\caption{Plots of a fixed element of the probability vector resulting from repeated executions of the one-layer tUPS circuit (with optimal, non-zero parameters) for $\ce{H2O}$ (4,4) as a function of time. The time is relative to when the corresponding test began. The index of the element is $51$ in the blocked spin-ordering convention, corresponding to the Hartree-Fock state. Each probability vector was obtained with $10^5$ shots. Both quantum hardware tests were done on \texttt{ibm\_fez}. The exact (noiseless) value of the fixed element is approximately $0.75$.}
\label{drift_stuff}
\end{figure}

Fig.~\ref{drift_stuff} shows how a fixed element of the unmitigated probability vector resulting from executions of the 1-layer tUPS circuit for $\ce{H2O}$ (4,4) can change over time on a quantum computer and noisy simulator. The index of the element is $51$ in the blocked spin-ordering convention, i.e. it corresponds to the bitstring $00110011$ which is the Hartree-Fock state when the bits are numbered starting from $0$ at the rightmost bit (as is the case here). The quantum computer tests were done on IBM's \texttt{ibm\_fez} backend by executing $150$ tUPS circuits in immediate succession as separate jobs with $10^5$ shots each. The same simulation test with a noise model imported from \texttt{ibm\_fez} is shown for reference.

\begin{figure}
\centering
  \centering
  \includegraphics[width=1\linewidth]{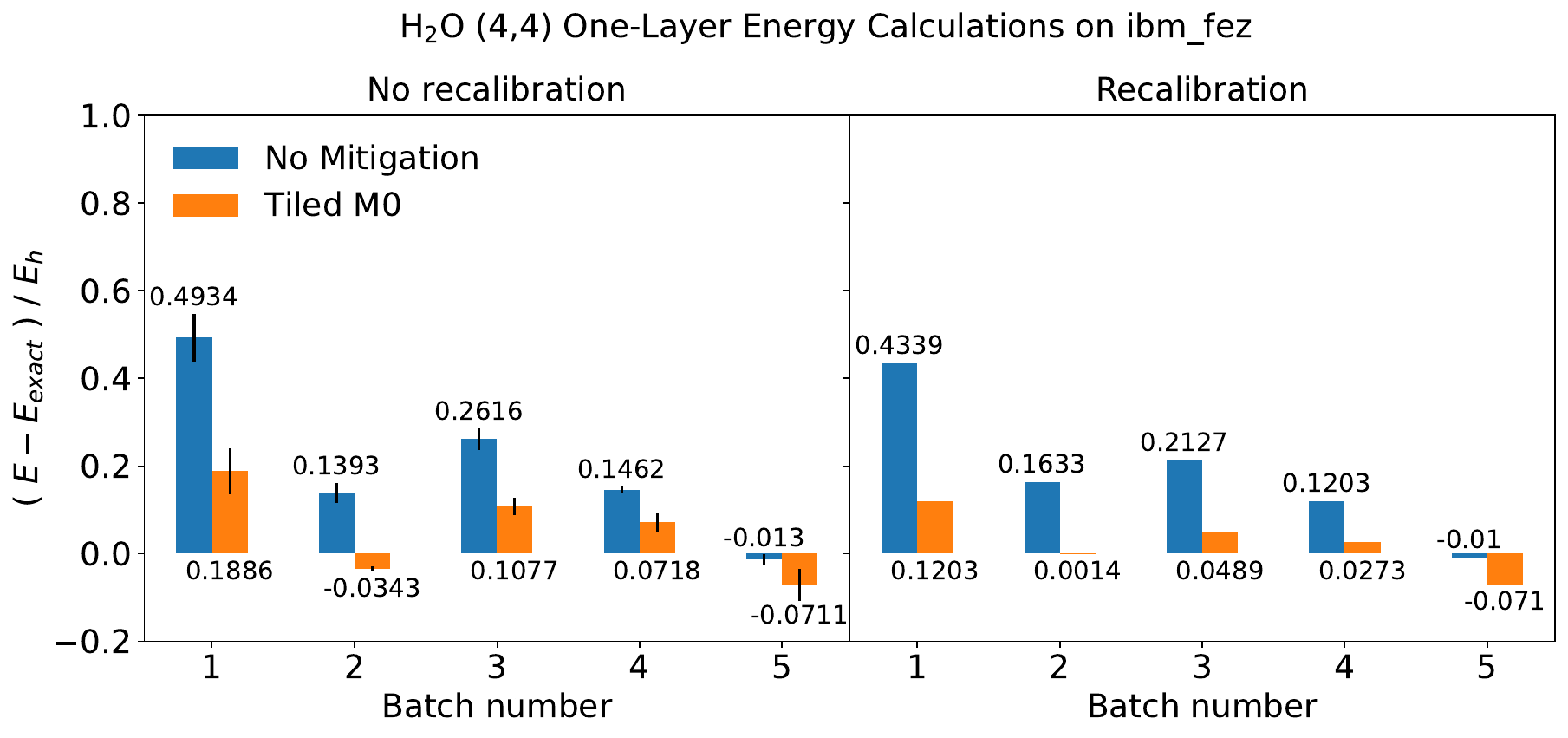}
\caption{Results from the recalibration experiment on \ce{H2O} (right) with batch energies from the original experiments (left) for reference. The standard deviations of the no-recalibration, unmitigated energies are $0.0546$, $0.0228$, $0.0256$, $0.0090$, and $0.0109 \, \mathrm{E_h}$ for batches $1$ to $5$, in that order (shown as error bars). The standard deviations of the tiled M0 energies are $0.0523$, $0.0049$, $0.0191$, $0.0206$ and $0.0371 \, \mathrm{E_h}$, in the same order.}
\label{fig:recalibration}
\end{figure}

In Fig.~\ref{fig:recalibration}, we show results from the \ce{H2O} confusion matrix recalibration experiment that we did where we partitioned the Hamiltonian into five smaller batches. For each batch, we calculated the energy as normal but with new noise characterization and confusion matrix construction. The figure shows the raw and error mitigated energy errors where the comparisons are with respect to the ideal energies for each batch. For reference, the left part of the figure shows average results from the experiments in Fig.~\ref{fig:h2o_experiments}. Note that no distinction between batches was made in the original, no-recalibration experiments, and the same confusion matrices were used for all batches. The right part of the figure shows the new results. The total number of energy measurement shots used for each batch in the recalibration experiment were between $2.6$ and $3.4$ million (see Table~\ref{batches_tabel}). The number of shots used for noise characterization for each batch was the same as in all other tests ($958,656$ in total). The batch numbers do not necessarily reflect the order in which the relevant quantum circuits were executed in the original experiments, i.e. batch number $1$ does not necessarily contain the Pauli strings whose quantum circuits were executed first in the original experiments. The batches were made manually based on the QWC-groups from the original experiments in such a way that no QWC-groups were split up. The condition numbers of the approximated full-system matrices in the recalibration tests are $3.19$, $3.24$, $3.44$, $2.96$ and $2.93$ for batches $1$ to $5$, in that order. Details about the operator batches are given in Table~\ref{batches_tabel} as well as the exact energies for each batch that we compare the raw and error mitigated energies with (excluding the contribution from the all-identity Pauli string). We note that the raw energy for batch number $5$ is actually better than the tiled M0 energy.

\begin{table}[t]
\begin{threeparttable}
\centering\renewcommand\cellalign{lc}
\setcellgapes{3pt}\makegapedcells
\caption{Details about the operator batches in the \ce{H2O} recalibration experiment.}
\centering\renewcommand\cellalign{lcc}
\setcellgapes{3pt}\makegapedcells
\begin{tabularx}{\linewidth}{l|XXXXXX}
\hline
Batch number & 1  &  2  &  3 & 4 & 5 \\  \hline  \hline
Pauli strings & 54 &  53 &  47 & 108 & 99 \\ 
Cliques &  13  & 17 & 9 & 16 & 18 \\ 
L1 norm & 2.16 &  1.75 &  1.65 & 1.84 & 2.04 \\
Shots, energy & $3,427,633$ & $2,775,274$ & $2,634,870$ & $2,918,547$ & $3,243,676$ \\
$E_{\mathrm{exact}} / \mathrm{E_h}$ & $-0.8453$ & $-0.4886$ & $-0.6675$ & $-0.2250$ & $0.0843$ \\

\bottomrule
\end{tabularx}

 \label{batches_tabel}
\end{threeparttable}
\end{table}
\section{Tiled M0 under excessive noise} \label{sec:excessive_noise}
The levels of hardware noise on a given backend can vary with time. If the noise is excessive, tiled M0 will fail. In Table~\ref{s1_meget_stoej}, we show results from calculations that we did on \texttt{ibm\_fez} where the noise was too severe for tiled M0 to succeed. The setups were the exact same in terms of shot counts etc. as for all the other experiments. The energies shown are from calculations with one layer in the tUPS Ansatz. Note that the condition numbers (shown in the last column of the table) are orders of magnitude larger than those reported in the figures in Sec.~\ref{sec:energy_calculations}, highlighting the stronger levels of noise that were present during the calculations.

\begin{table}[t]
\begin{threeparttable}
\centering\renewcommand\cellalign{lc}
\setcellgapes{3pt}\makegapedcells
\caption{Results from quantum experiments on \texttt{ibm\_fez} where tiled M0 failed completely because of severe noise. The energies were all calculated with one layer in the tUPS Ansatz. The condition numbers for the approximated full-system confusion matrices are shown in the last column.}
\centering\renewcommand\cellalign{lcc}
\setcellgapes{3pt}\makegapedcells
\begin{tabularx}{\linewidth}{lXX|X}
\hline
$(E - E_{exact}) \, / \, E_h$ with: & Tiled M0  &  No Mitigation & Condition number \\  \hline
Butadiene experiment & $250.65$ & $0.62004$ & $79,431$ \\ 
Butadiene experiment & $-3.18234$ & $0.66804$ & $458.2$\\
\ce{H2} experiment & $0.282205$ & $0.564870$ & $39.3$ \\

\bottomrule
\end{tabularx}
 \label{s1_meget_stoej}
 
\end{threeparttable}
\end{table}